\documentclass[compsoc,conference,a4paper,10pt,times]{IEEEtran}
\usepackage[affil-it]{authblk}
\usepackage{amsmath}
\usepackage{amssymb}
\usepackage{array}
\usepackage{balance}
\usepackage{booktabs}
\usepackage[font={small}]{caption}
\usepackage{enumitem}
\usepackage[multiple]{footmisc}
\usepackage{graphicx}
\usepackage[misc]{ifsym}
\usepackage{makecell}
\usepackage{multirow}
\usepackage[font={small}]{subcaption}
\usepackage{times}
\usepackage{titlesec}
\usepackage{xcolor}

\makeatletter
\def\@citex[#1]#2{\leavevmode
\let\@citea\@empty
\@cite{\@for\@citeb:=#2\do
{\@citea\def\@citea{,\penalty\@m\ }%
\edef\@citeb{\expandafter\@firstofone\@citeb\@empty}%
\if@filesw\immediate\write\@auxout{\string\citation{\@citeb}}\fi
\@ifundefined{b@\@citeb}{\hbox{\reset@font\bfseries ?}%
\G@refundefinedtrue
\@latex@warning
{Citation `\@citeb' on page \thepage \space undefined}}%
{\@cite@ofmt{\csname b@\@citeb\endcsname}}}}{#1}}
\makeatother

\def\tablename{Table}

\begin{document}

\title{WatchAuth: User Authentication and Intent Recognition\\in Mobile Payments using a Smartwatch}
\author{Jack Sturgess, Simon Eberz, Ivo Sluganovic, and Ivan Martinovic}
\affil{Department of Computer Science, University of Oxford, Oxford, UK \\\{firstname.lastname\}@cs.ox.ac.uk}
\maketitle

\begin{abstract}
In this paper, we show that the tap gesture, performed when a user `taps' a smartwatch onto an NFC-enabled terminal to make a payment, is a biometric capable of implicitly authenticating the user and simultaneously recognising intent-to-pay. The proposed system can be deployed purely in software on the watch without requiring updates to payment terminals. It is agnostic to terminal type and position and the intent recognition portion does not require any training data from the user. To validate the system, we conduct a user study (n=16) to collect wrist motion data from users as they interact with payment terminals and to collect long-term data from a subset of them (n=9) as they perform daily activities. Based on this data, we identify optimum gesture parameters and develop authentication and intent recognition models, for which we achieve EERs of 0.08 and 0.04, respectively.
\end{abstract}


\section{Introduction}

The popularity of cashless and contactless payment systems continues to grow. In the UK, card payments surpassed cash payments for the first time in 2018 \cite{MobileTransaction2020}. NFC-enabled mobile payments are projected to account for over 27\% of market share by the end of 2021 \cite{Technavio2020}, driven by user preference for usability and enhanced security \cite{Huh2017}.

Mobile payment systems (also known as tap-and-pay), such as Google Pay, enable the user to provision one or more payment cards to a virtual wallet that is accessible via a smartphone, which can then be used to make payments over NFC (using tokenisation to protect card details). Typically, these systems require two factors to authenticate the user and utilise the fingerprint reader or face recognition camera of the device where possible.

In recent years, mobile payment systems have had their functionality extended onto smartwatches. When paired with a smartphone, a watch can access the same virtual wallet and then make payments over NFC, even when the phone is not present. Current smartwatches do not offer fingerprint readers or cameras, but use short passcodes for authentication. For convenience, a smartwatch can be configured to remain unlocked (\textit{i.e.}, in an authenticated state) after a single passcode entry until it is restarted or removed from the wrist. Apple Pay Express Travel mode enables payments to be made by an Apple phone or watch at busy transport barriers without it needing to be unlocked, but limiting the scope of misuse by working only for certain merchant codes. In each case, explicit authentication actions are obviated to improve usability at a potential cost to security. Moreover, without explicit user interaction with the terminal, an uncertainty arises for the payment provider as to whether the user intended to make the payment at all.

In 2020, the European Union introduced the Updated Payment Services Directive (PSD2) \cite{Thales2020}, overhauling payment regulations for the banking sector and establishing a precedent that other nations are likely to follow. One of its core principles, entitled Strong Customer Authentication, mandates the use of multi-factor authentication for all payment transactions. To fulfil this, a system must verify the user's identity with at least two independent factors that are based on either \textit{knowledge} (something only the user knows), \textit{possession} (something only the user possesses), or \textit{inherence} (something only the user is). For mobile payments, possession of the smart device (and its tokens) counts as one factor, so at least one more is needed.

Given the trend towards convenience, implicit factors that do not require any user effort are becoming more desirable. Fuelled by the availability of multifarious embedded sensors, recent work  has demonstrated the use of a variety of factors, such as GPS location, usage habits, environmental conditions, proximity to other devices, and behavioural biometrics to authenticate the user continuously and unobtrusively on Web browsers \cite{Freeman2016}, smartphones \cite{Gupta2012,Hayashi2013,Hocking2013,Jakobsson2009,Kayacik2014,Lee2016,Miettine2014,Riva2012,Yao2017}, and wearable devices \cite{Mare2014,Mare2019,Yang2015}. Shrestha \textit{et al.} \cite{Shrestha2016} showed that users making tap-and-pay payments with a smartphone can be authenticated by their tap gestures using various sensors.

In this work, we focus on implicit authentication and intent recognition in mobile payments using a smartwatch. We are motivated firstly by the lack of biometric authentication options that are available to smartwatch users and secondly by usability advancements that may cast doubt on user intent during the payment process. We conduct a user study to collect wrist motion data from users as they interact with point-of-sale terminals and show that the tap gesture is sufficiently distinct between users that it can be used to authenticate users. We also collect a large dataset of relevant non-tap gestures from users as they perform other activities and show that the tap gesture is sufficiently recognisable between gestures that it can be used to infer whether a payment is intentional, providing a new technique to allay uncertainty in convenience-optimised payment schemes and to strengthen system security by rejecting unintentional payments.

\textbf{Contributions.}
\begin{itemize}
\vspace{-\topsep}
\item Using only wrist motion data, we show that a single tap gesture performed by the user while making a payment with a smartwatch can authenticate the user and recognise intent-to-pay, both implicitly and simulateously. Our system runs on the watch and does not require any changes to terminals.
\item We show that our approach can be applied to real-time data for in-store usage and to historic data for retrospective fraud detection, offering a defence against malicious, shared, and accidental payments.
\item Our authentication model is terminal-agnostic, so does not need any specific terminal type or position.
\item Our intent recognition model is user-agnostic, so does not need training data from the user during the enrolment phase and is innately resistant to drift.
\item We implement our system to allow for real-world evaluation and show that our models can be tuned for use as a second factor in an existing system, providing a strict improvement to security (by adding a layer of false acceptance detection and introducing an unsharable factor) with negligible cost to usability.
\item We make the code and data required to reproduce our results available at http://github.com/jacksturgess.
\end{itemize}

\textbf{Paper Structure.} The rest of this paper is organised in the following way. Section \ref{sec:Background} presents a summary of gesture biometrics. Section \ref{sec:ObjectivesandAssumptions} outlines the challenges of using a smartwatch compared with a smartphone and details our system and threat models. Section \ref{sec:ExperimentalDesign} describes our experimental apparatus. Section \ref{sec:Methods} explains the methods\\that we employ to collect and process data, train classifiers, and measure performance. Section \ref{sec:Results} presents and analyses our results. Section \ref{sec:Discussion} discusses peripheral topics. Section \ref{sec:RelatedWork} compares our approach to related work. Section \ref{sec:LimitationsandFutureWork} considers limitations. Section \ref{sec:Conclusion} concludes the paper.

\section{Background}\label{sec:Background}

In the context of authentication, a biometric is a measured characteristic of an individual that should be unique, persistent, and hard to impersonate. A biometric can be categorised as either physiological or behavioural. Physiological biometrics measure a physical feature of the user, such as a fingerprint or retina scan. Behavioural biometrics measure patterns of movements exhibited by the user, such as gait or keystroke dynamics. Biometric authentication systems first require an enrolment phase, in which features from the user's biometric measurements are extracted and encapsulated in a template; when the user attempts to authenticate, a fresh measurement is taken and features are extracted anew and matched against the template. Behavioural biometrics tend to need a longer enrolment phase, which had rendered them largely impractical before ubiquitous sensors became available.

Physiological biometrics are typically measured in discrete, explicit actions performed solely for the purpose of authentication, such as touching or looking at a sensor. Behavioural biometrics are measured over time and without any effort on the part of the user---such factors are referred to as \textit{implicit}, because they can be measured while the user engages in other tasks. Implicit factors facilitate \textit{continuous} authentication, where a system authenticates the user often and unobtrusively to maximise its confidence in his identity at all times.

In terms of PSD2, all biometrics are inherence-based. Biometrics cannot be forgotten or guessed (like knowledge-based factors) or lost or stolen (like possession-based factors), but they can \textit{drift} (naturally change) over time. Lee \textit{et al.} \cite{Lee2017} demonstrated a mitigation technique for drift using an update mechanism to replace the old user template with a new one by averaging it with the latest signal. User-agnostic systems are resistant to drift as they are not trained on the individual user, so changes in his behaviour are irrelevant. A template could also be \textit{poisoned} (maliciously changed) by exploiting the update mechanism such that the template is gradually morphed to wrongly accept an impersonator's signals \cite{Biggio2013,Lovisotto2020}, although this too can be mitigated, here by limiting the frequency of updates or gating the mechanism behind another authentication factor. Biometrics cannot be shared between users, ensuring a one-to-one identity mapping, but mimicry attacks are typically feasible given sufficient resources and trait collisions can occur in large user sets.

A gesture is a series of movements made by the user; it could be explicit, such as a nod to indicate an affirmative response or a touchscreen swiping pattern to activate some functionality (\textit{e.g.}, the \textit{pinch gesture} to resize or zoom), or it could be performed innately as part of another activity and thus be implicit, such as the movement of the wrist while typing. A gesture as a biometric is typically measured using inertial sensors, such as accelerometers and gyroscopes. Early work in gesture-based authentication focused on explicit gestures, showing that users could be distinguished from one another based on their performing an explicit action, such as various arm-swinging gestures, with handheld devices \cite{Liu2009,Matsuo2007,Okumura2006} or using wrist-worn sensors \cite{Liang2017,Yang2015,Yu2020}. More recent work has considered implicit gestures and their feasibility for use in continuous authentication. Frank \textit{et al.} \cite{Frank2013} use sensors in a smartphone to authenticate the user based on touchscreen interaction over time. Han \textit{et al.} \cite{Han2018} use sensors mounted on smart home devices to identify occupants based on object interaction. Nassi \textit{et al.} \cite{Nassi2016} use wrist-worn sensors to verify users as they hand-write signatures and Griswold-Steiner \textit{et al.} \cite{Griswold-Steiner2017} extend the idea to general handwriting.

One drawback of using wrist-worn sensors is that users tend to perform pertinent gestures with their dominant hand and wear a smartwatch on the wrist of the other. In our case, this problem is avoided as the user must move the watch itself to the terminal when making a payment.

\section{Objectives and Assumptions}\label{sec:ObjectivesandAssumptions}

\subsection{Design Considerations}\label{sec:DesignConsiderations}

We have chosen to focus on the use of a smartwatch, rather than a smartphone, due to the following two challenges that make it a more interesting problem.

Firstly, the starting point of a tap gesture is more difficult to determine on a watch. A phone is picked up with an explicit gesture, providing an indicator, whereas a watch is worn continuously. While the continuous collection of motion data has only a negligible impact on the battery life of our watch, gesture segmentation and classification tasks have a greater impact and cannot practically be done continuously with current hardware. At present, this problem is avoided because an explicit action is required on all devices to initiate a payment (\textit{e.g.}, double-clicking a side button); however, in convenience-optimised, zero-interaction payment schemes, this parity breaks. To ensure that our data reflects the most difficult scenario, we preclude participants in our user study from interacting with the watch between tap gestures. We address the challenge of finding starting points by representing tap gestures with sliding time windows of sensor data extended backwards from the NFC contact point.

Secondly, a watch undergoes a much greater change in orientation to perform a tap gesture. A phone can be tapped against the terminal without any change in orientation, owing to its sensor placement and the user's arm not being in the way; whereas, a watch is constrained to the wrist of the user and so must follow the physiology of the arm as it is moved until the watch face rests against the terminal. Moreover, the sensor axes are relative to the orientation of the device (see Section \ref{sec:ExperimentSensorModule}); thus, although the watch travels from the user to the terminal along a single axis in the external reference frame, its movement is measured in the reference frame of the watch along multiple sensor axes as it changes orientation during travel. One way in which to stabilise this behaviour would be to transform the sensor data into the external co-ordinate system (\textit{e.g.}, as done by Ard{\"u}ser \textit{et al.} \cite{Arduser2016}). Unfortunately, this would require some ground truth from both reference frames to calibrate the transformation, which is not practical in our case---either we would need (i) to prompt the user to hold the watch in a known position, which would impose an inconvenience, or (ii) to infer backwards from when the watch face is flush against the terminal, which would require us to know the position of the terminal. We instead address this challenge by extracting a number of axis-invariant features to inform our classifiers.

\subsection{System Model}\label{sec:SystemModel}

We consider a system model in which a user is wearing a smartwatch on his wrist and is using it to make NFC-enabled payments at point-of-sale terminals in a typical setting---namely, in a shop or at the entry barriers to a transport system. To make a payment, we assume that the user performs a \textit{tap gesture} by moving his wrist towards the terminal until the watch is near enough to exchange data via NFC. The NFC contact point is when the payment provider would decide whether to approve the payment, so we assume that this marks the end of the tap gesture for real-time purposes. We assume that the watch has an accelerometer and gyroscope and that we have access to the data generated by these sensors. We use data from the inertial sensors only, regardless of the availability of any medical, environmental, or location sensors that the watch may have; this enables us to compare our results fairly with those of related works (in Section \ref{sec:RelatedWork}).

While the current generation of smartwatches are dependent upon a paired smartphone for administration purposes and the installation of apps, they may operate independently once set up. As such, we do not assume or require that a phone be present and we do not make use of any additional data, such as location or proximity data, that one might provide.

Using the wrist motion data collected from the smartwatch, we create two separate models: an \textit{authentication model}, in which we verify the identity of the user, and an \textit{intent recognition model}, in which we infer the intention of the user to make the payment. For the latter, we assume that a tap gesture is composed of a sufficiently obscure, deliberate sequence of movements as to be unlikely to be performed unintentionally, such that if we identify a tap gesture during a transaction then we infer that the payment is intentional. The combination of these models forms our system, which we call WatchAuth.

\begin{figure}[t!]
	\centering
  	\includegraphics[width=0.68\linewidth]{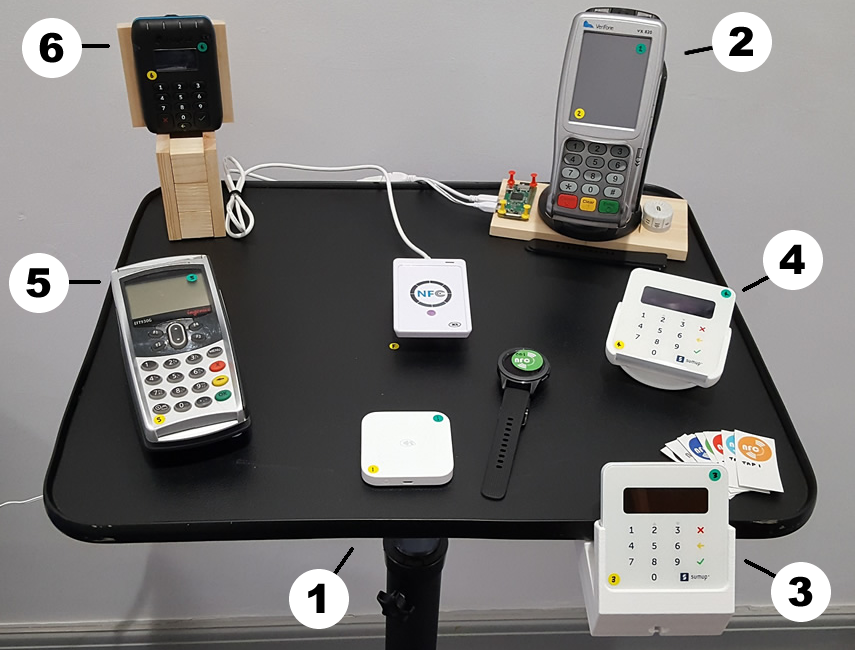}
   	\caption{The equipment used in our experiment: six fixed terminals (labelled, see Table \ref{tab:TerminalDetails} for details), an NFC reader, a Raspberry Pi for timestamp collection, and a smartwatch.}
   	\label{fig:ExperimentOverview}
\end{figure}

\begin{table}
	\centering
	\small
	\begin{tabular}{c|c|c|c}
		\toprule
		\textbf{Terminal} & \textbf{Height (cm)} & \textbf{Tilt ($^{\circ}$)} & \textbf{Distance (cm)} \\
		\midrule
		1 & 100 & 0 & 5 \\
		2 & 120 & 60 & 25 \\
		3 & 95 & 45 & -10 \\
		4 & 105 & 30 & 15 \\
		5 & 110 & 15 & 10 \\
		6 & 115 & 90 & 30 \\
		\midrule
		F & \multicolumn{3}{c}{picked up from centre of platform} \\
		\bottomrule
	\end{tabular}
	\caption{Details of the terminals used in our experiment; the indices match those labelled in Figure \ref{fig:ExperimentOverview} and `F' is the freestyle terminal. \textit{Height} is measured from the floor to the lowest point of the terminal; \textit{Tilt} is the inclination at the lowest point of the terminal; and \textit{Distance} is measured from the front of the stand to the point of the terminal that is closest to the user. Terminals 2 and 6 match terminals on self-service checkouts at supermarket chains, roughly an arm's length from the user.}
	\label{tab:TerminalDetails}
\end{table}

The principal goal of this work is to show that the tap gesture is a biometric capable of authenticating the user and recognising intent-to-pay, implicitly and simultaneously. To demonstrate the applicability of our approach, we evaluate our models against our threat model in three use-cases. Firstly, for \textit{in-store usage}, we restrict ourselves to using sensor data that is available in real-time at the terminal, collected before the NFC contact point, pursuant to the assumption given above. Secondly, for \textit{retrospective fraud detection}, we assume that a payment provider has access to historic sensor data, collected before and after the NFC contact point, and we extend our models accordingly. Thirdly, for use as an \textit{additional factor}, we optimise our models to minimise the occurrence of false negatives, so that implementing our models alongside an existing authentication system provides only a strict improvement to security with negligible cost to usability.

\begin{figure}[t!]
	\centering
  	\includegraphics[width=0.84\linewidth]{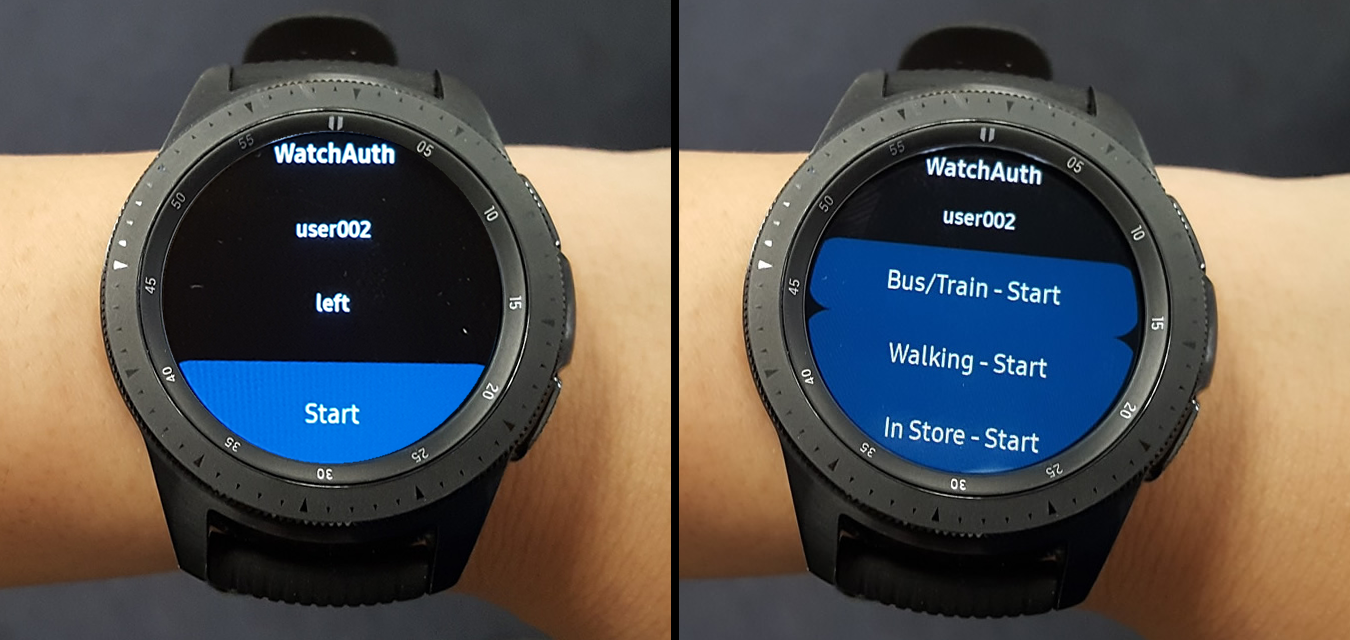}
   	\caption{The two modes of our data collection app, for in-lab (left) and out-of-lab (right) usage.}
   	\label{fig:ExperimentApps}
\end{figure}

\begin{figure}
	\centering
  	\includegraphics[width=0.84\linewidth]{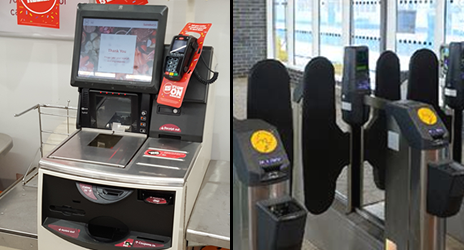}
   	\caption{Two examples of point-of-sale terminals commonly found in the UK: the terminal on a self-service checkout at a supermarket (left) and the terminals on a train station barrier (right, in yellow). The positions of these terminals are replicated by our Terminals 2 and 3, respectively.}
   	\label{fig:ExperimentTerminals}
\end{figure}

\subsection{Threat Model}\label{sec:ThreatModel}

Given that our system model branches into two, our threat model considers a separate attacker against each.

For the authentication model, we consider an adversary that has possession of a legitimate user's smartwatch, has unlocked it, and is attempting to use it to make a payment at an unstaffed terminal. The adversary may have (maliciously) stolen the watch or (benignly) borrowed it; we include the latter as we seek to prevent the user from sharing the watch for payment purposes. Our goal here is to authenticate the legitimate user and to reject other users by using only tap gestures.

For the intent recognition model, we consider that the user has unlocked the smartwatch and that, while it remains on the wrist and in an unlocked state, an unintentional payment has been initiated. This could be malicious, where an adversary may have moved a terminal or other NFC-enabled device to the watch unbeknownst to the user, such as a skimming attack, or it could be accidental. We assume that the user is wearing the watch and performing nondescript activities in any of three public settings: while (i) \textit{walking} or (ii) commuting \textit{on a bus or train}, where an adversary would have ample access to the watch, or (iii) \textit{in a shop}, where an accidental payment may be mistaken for intentional because of its location. Our goal here is to recognise that the user did not perform a tap gesture at the time of the payment and therefore did not intend to make the payment.

In this work, we concentrate on the extent to which gesture biometrics can be used to defend against these attacks. We do not consider threats to other components in the payment system, tampering of devices or biometric templates, malware, or denial of service attacks.

\section{Experimental Design}\label{sec:ExperimentalDesign}

\subsection{Experiment Overview}

To evaluate the extent to which wrist motion data can be used to achieve our goals, we designed and conducted a user study to collect data. Our experiment consists of six point-of-sale terminals on an adjustable stand fixed at a height of 100 cm, an ACR122U NFC reader connected to a Raspberry Pi, and a Samsung Galaxy Watch running the Tizen 4.0 operating system (as shown in Figure \ref{fig:ExperimentOverview}). The experiment also includes a screen, connected wirelessly to the Raspberry Pi, that instructs the user when to perform each tap gesture.

We built a data collection app with the Tizen Studio IDE and installed it on the smartwatch to continuously collect timestamped wrist motion data as the user wears it (as shown in Figure \ref{fig:ExperimentApps}). To collect data for a single tap gesture, the NFC reader is first affixed to the front of the terminal and the user performs the tap gesture on it as if making an NFC-enabled payment. Each NFC contact point timestamp is captured by the Raspberry Pi and a short spacing delay is initiated before the user is instructed to perform the next tap gesture.

A subset of participants also collected data outside of the lab by wearing the smartwatch during various non-payment activities. The user selects the setting and the app collects and labels the motion data until stopped.

\subsection{Point-of-Sale Terminals}

To emulate real-world mobile payment scenarios, we capture tap gestures using \textit{seven} terminals: six in fixed positions (as shown in Figure \ref{fig:ExperimentOverview} and detailed in Table \ref{tab:TerminalDetails}) and one `freestyle'. For five of the six fixed terminals, we surveyed prominent supermarket and restaurant chains to find popular or standardised terminal positions (in terms of height, tilt angle, and distance from the user) and set our terminals to match common configurations. We set the other fixed terminal to match the position of the terminal on a train station barrier (as shown in Figure \ref{fig:ExperimentTerminals}). For the freestyle terminal, the user picks up the NFC reader directly with his other hand and performs a tap gesture against it, returning it after each interaction, just as if a vendor had handed an unmounted terminal to a customer. We chose these scenarios to represent a broad cross-section of the real-world instances in which a smartwatch user may be required to perform a tap gesture as part of a payment transaction.

The six fixed terminals remain deactivated throughout the experiment as their functionality is not required. For consistent data collection, we affix the NFC reader to the front of each terminal when using it. As such, the terminals should be regarded only as fixtures that enforce heights and angles, as well as a tool for immersing the user in a payment scenario.

During the user study, each participant is instructed to stand in front of the terminal platform while performing his tap gestures. Aside from this, we do not prescribe any constraints on positioning as we want the user to stand and interact with the terminals comfortably and naturally as though making payments in a real-world setting.

\subsection{Sensor Module}\label{sec:ExperimentSensorModule}

The Tizen platform provides four inertial sensors directly or derived from the MEMS accelerometer and gyroscope in the smartwatch. The \textit{accelerometer} measures change in velocity and the \textit{gyroscope} measures angular velocity. The \textit{linear accelerometer} is derived from the accelerometer with the effects of gravity excluded. The \textit{gyroscope rotation vector} (GRV) is a fusion of accelerometer and gyroscope readings to compute the orientation of the device. Our app collects timestamped data from all four at a sampling rate of 50 Hz.

\begin{figure}[t!]
	\centering
  	\includegraphics[width=0.84\linewidth]{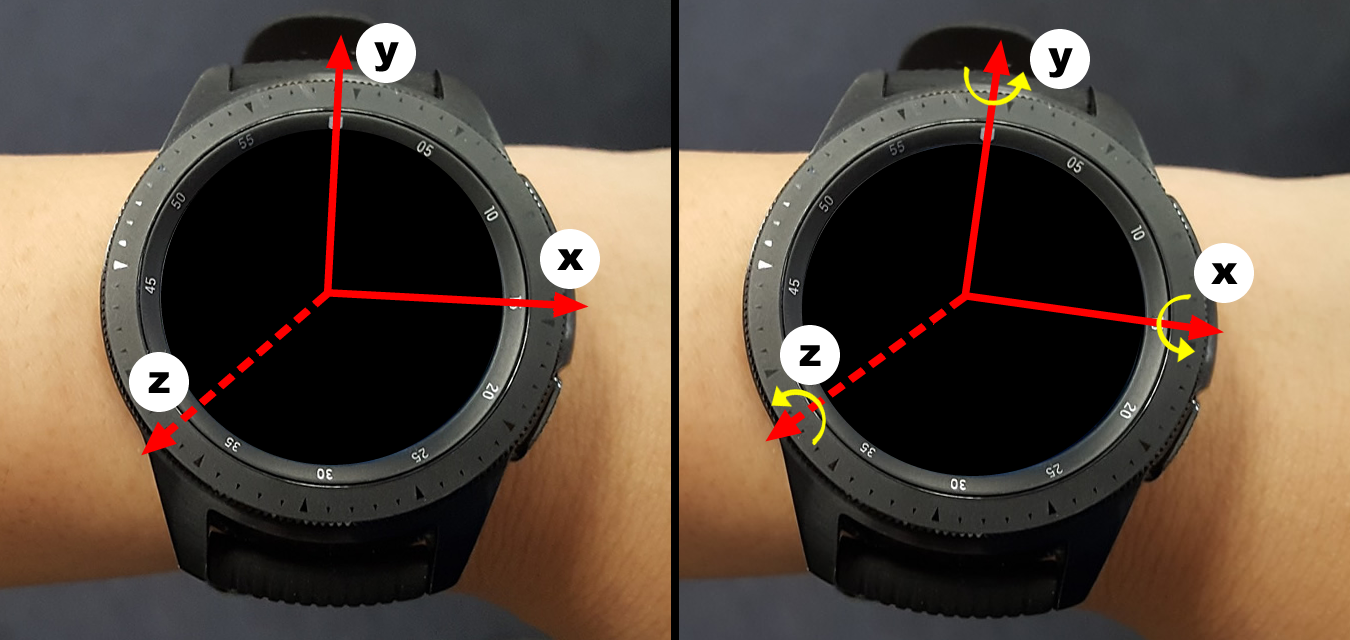}
   	\caption{The sensor axes relative to the frame of the smartwatch; the positive $x$-axis points towards the hand when worn on the left wrist; angular velocity is measured along the yellow arrows.}
   	\label{fig:ExperimentSensors}
\end{figure}

The inertial sensors measure wrist motion along three axes that are relative to the frame of the watch (as shown in Figure \ref{fig:ExperimentSensors}). Motion along the $x$-axis corresponds with arm extension or withdrawal; the $y$-axis, with side-to-side arm waving; and the $z$-axis, directly up- and downwards through the watch face.

\subsection{User Study}

\begin{figure}
	\centering
  	\includegraphics[width=0.84\linewidth]{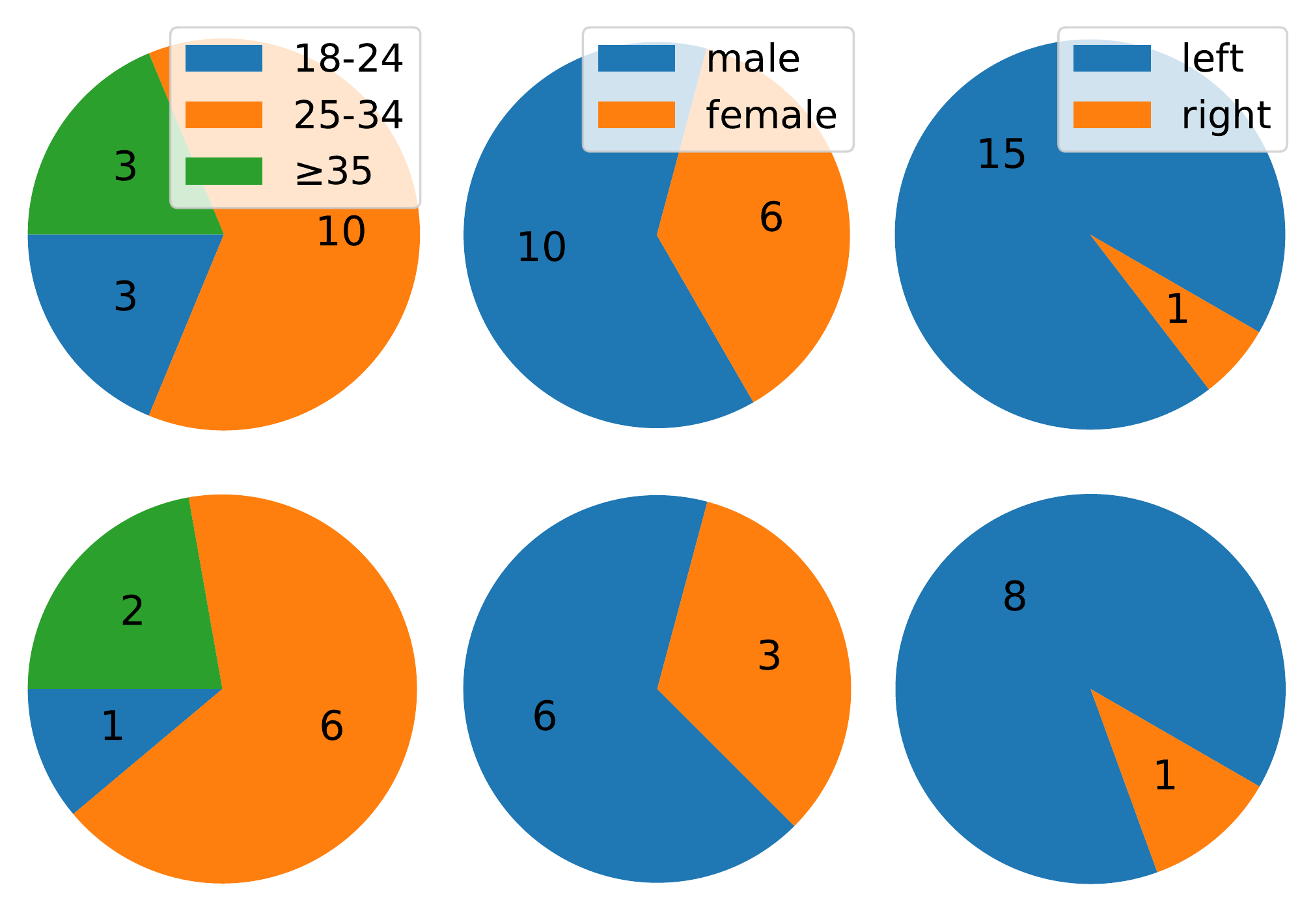}
   	\caption{The distribution of age (left), sex (centre), and on which wrist the smartwatch was worn (right) of users in our main lab experiment (top, n=16) and the subset of users that collected additional data (bottom, n=9).}
   	\label{fig:DataDemographics}
\end{figure}

To collect data, we conducted a user study that was reviewed and approved by the relevant research ethics committee at our university. We recruited 16 participants, including students and members of the public; a breakdown of participant demographics is shown in Figure \ref{fig:DataDemographics}.

Each participant attended 3 data collection sessions. In each session, the participant performed 10 tap gestures on each of the seven terminals. The first and second sessions occurred back-to-back, separated by a short break lasting roughly 5 minutes; the third session occurred on a different day. In total, we collected 210 tap gestures from each user.

A subset of 9 participants also collected and labelled data outside of the lab while walking, commuting on a bus or train, or in a shop---daily activities identified in our threat model as likely settings for an unintentional payment. In total, we collected 1,088 minutes of out-of-lab activity data across all users (601 minutes walking, 317 minutes on a bus or train, 148 minutes in-store, and 22 minutes of combined activities\footnote{We refer to an activity as \textit{combined} if the user performed multiple of these activities, each exclusively, but did not label them individually.}).

\textbf{User Statistics.} Of the participants, 19\% indicated that they regularly wore a non-smart watch and 38\% wore a smartwatch; only 6\% had ever made a payment using a smartwatch compared with 81\% who had paid using a smart\textit{phone}. The tap gesture is intuitive to perform; we observed no difference, either during the user study or in subsequent analysis, between those who regularly wore any watch and those who did not, nor between those who had paid with a smart device and those who had not.

\section{Methods}\label{sec:Methods}

\subsection{Data Processing}

We collect timestamped data from four inertial sensors. Each accelerometer, gyroscope, or linear accelerometer sample is given in the form $(t, x, y, z)$ and represents the change in velocity or angular velocity of the smartwatch along each axis at time $t$. Each GRV sample is given as a quarternion in the form $(t, x, y, z, w)$ and approximates the orientation at time $t$.

We express a tap gesture using series of inertial sensor data samples within a time window. To retrieve the tap gestures for each user, we segment 4-second blocks of sensor data by using the NFC contact point timestamps as the endpoint of each window. We found that a 4-second maximum window size was sufficient to encapsulate the entirety of each gesture.

\begin{figure*}[t!]
	\centering
	\begin{subfigure}[t]{0.37\textwidth}
		\centering
		\includegraphics[height=3.9cm]{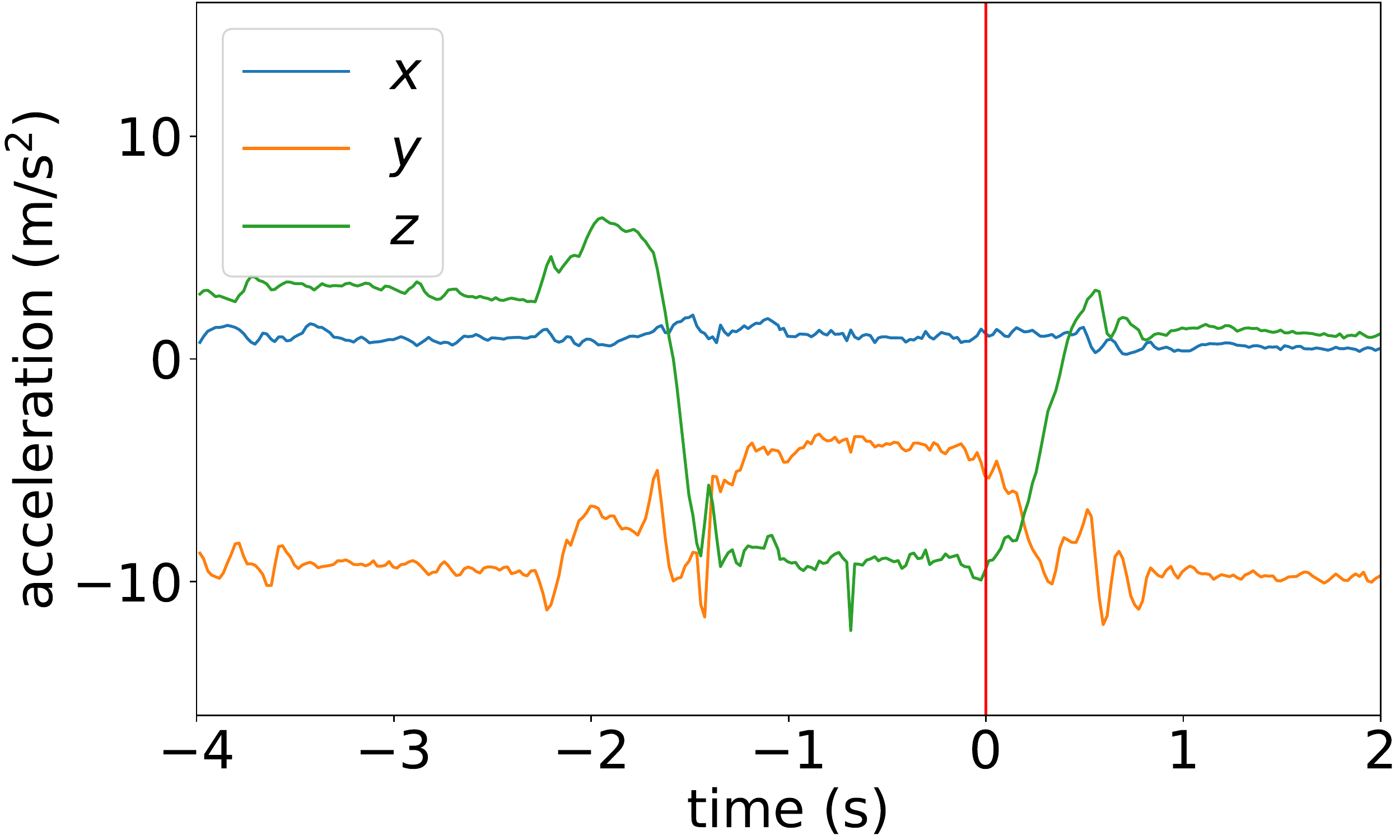}
		\caption{accelerometer data\textcolor{white}{\\.}}
		\label{fig:DataProcessingPlotAcc}
	\end{subfigure}
	~
	\begin{subfigure}[t]{0.41\textwidth}
		\centering
		\includegraphics[height=3.9cm]{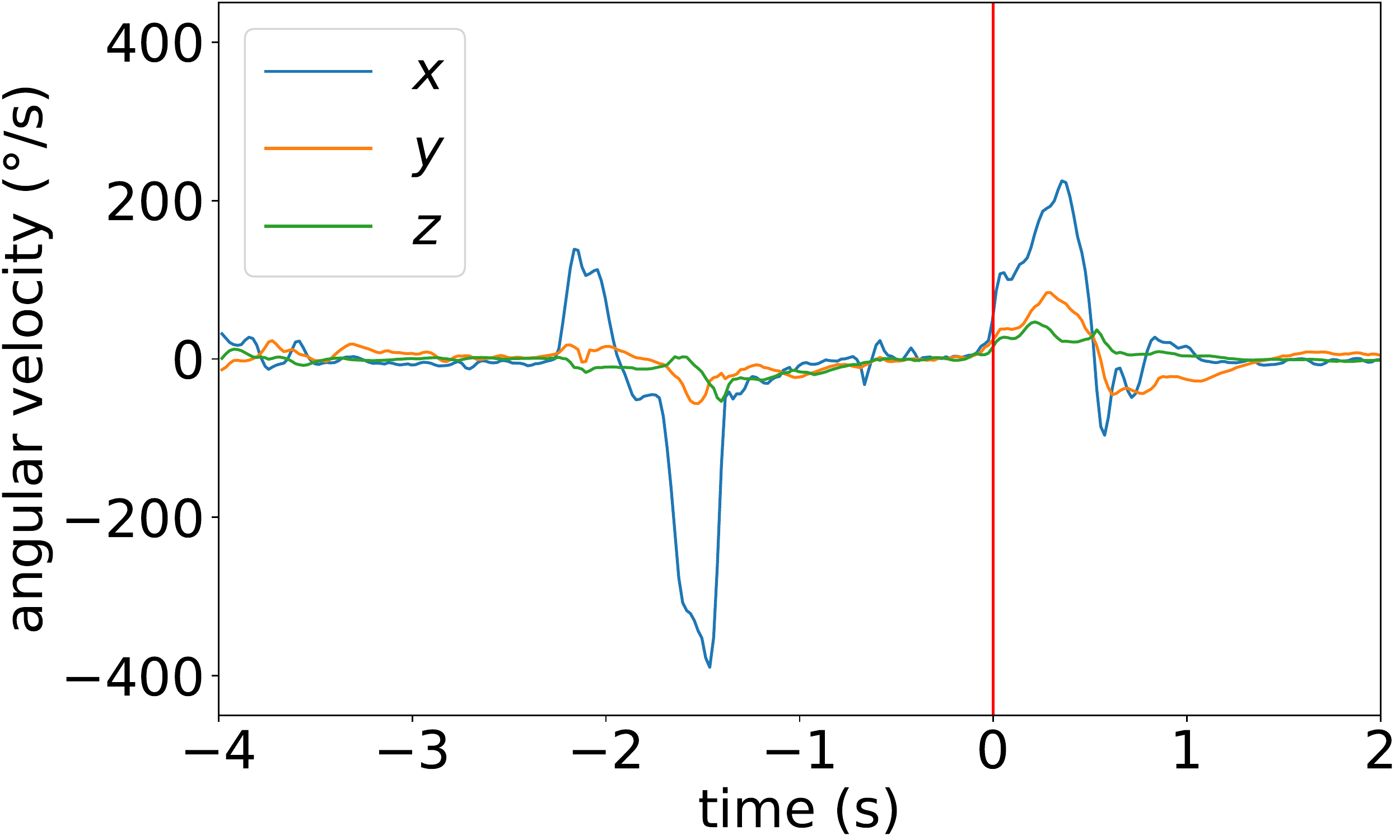}
		\caption{gyroscope data\textcolor{white}{\\.}}
		\label{fig:DataProcessingPlotGyr}
	\end{subfigure}
	~
	\begin{subfigure}[t]{0.37\textwidth}
		\centering
		\includegraphics[height=3.9cm]{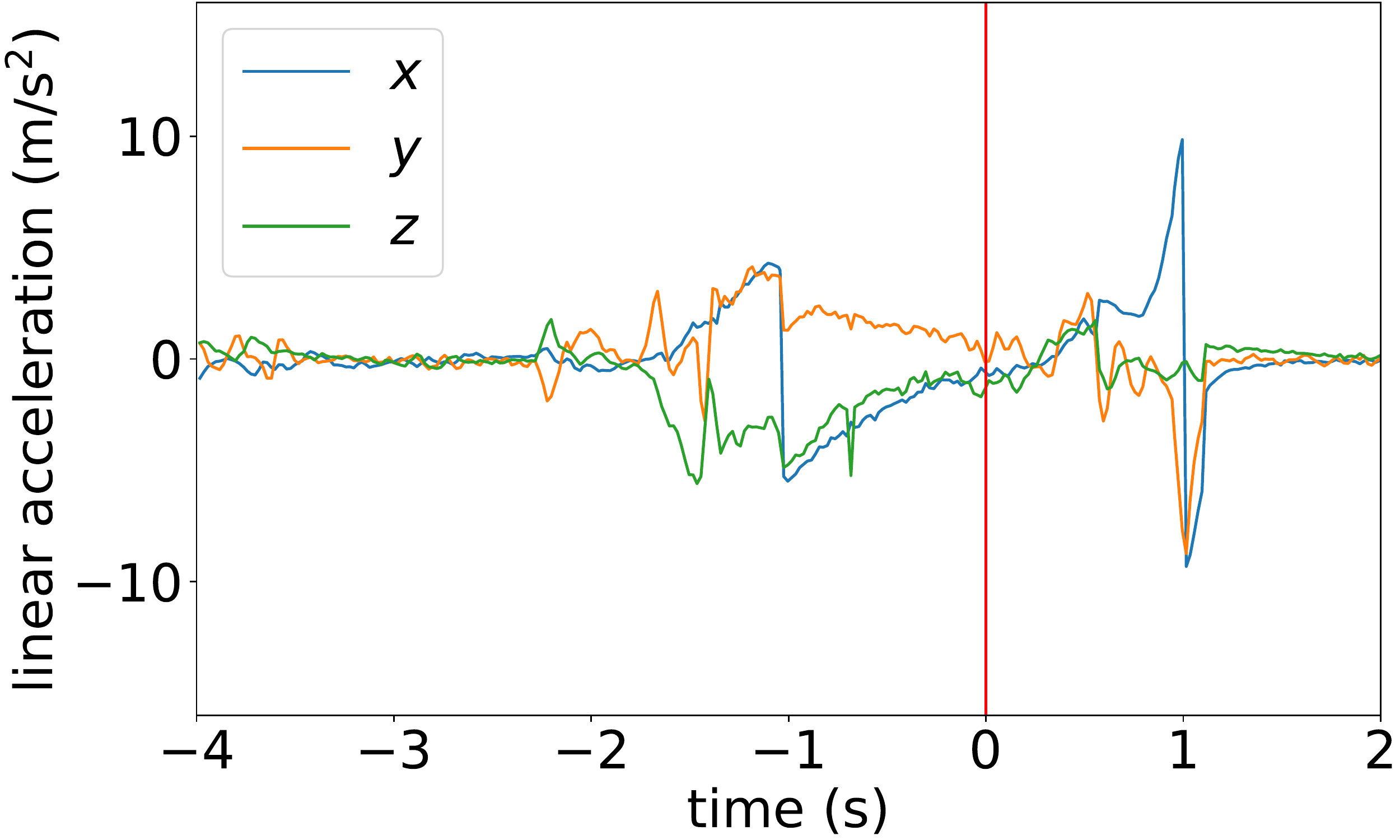}
		\caption{linear accelerometer data}
		\label{fig:DataProcessingPlotLAc}
	\end{subfigure}
	~
	\begin{subfigure}[t]{0.41\textwidth}
		\centering
		\includegraphics[height=3.9cm]{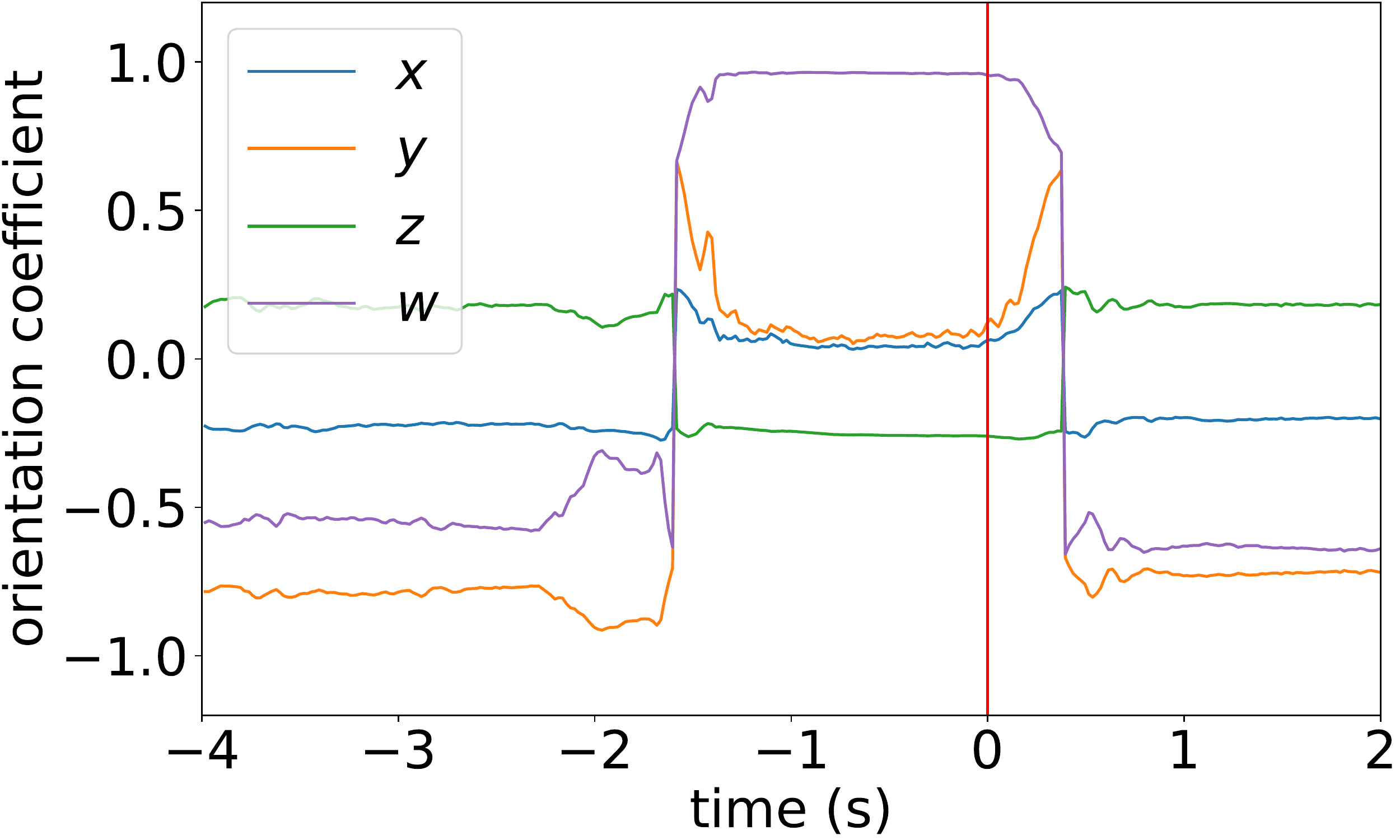}
		\caption{GRV data}
		\label{fig:DataProcessingPlotGRV}
	\end{subfigure}
	\caption{Visualisations of the same tap gesture, showing data 4 seconds before the NFC contact point and 2 seconds after from the four sensors used to collect our data. The NFC contact point is set at time 0 (indicated with a red line).}
	\label{fig:DataProcessingPlots}
\end{figure*}

Sensor data for an exemplar tap gesture is shown in Figure \ref{fig:DataProcessingPlots}. Here, we infer from the accelerometer data that the smartwatch reached the terminal at approximately 1.5 seconds before the NFC contact point and then, from the gyroscope and GRV data, that the user adjusted the orientation of the watch face to align it with the terminal to find the NFC connection. An additional 2 seconds of data after the NFC contact point is included for context, showing the user's arm withdrawing.

To investigate optimum tap gesture parameters, we compare (in Section \ref{sec:Results}) the performances of gestures bounded by various window sizes and offsets, where the offset is the time between the NFC contact point and the end of the window. For an NFC contact point timestamp $T_0$, a window size $s$, and an offset $o$, we retrieve a tap gesture with start time $T_S$ and end time $T_E$, where $T_E=T_0-o$ and $T_S=T_E-s$.

We segment the out-of-lab sensor data into comparable 4-second blocks. In total, we obtained 32,441 4-second, non-tap gestures by segmenting the data every 2 seconds to ensure a 50\% overlap so as not to eliminate any interesting regions.

\subsection{Feature Extraction}

Whenever a gesture with given parameters is retrieved, we apply a low pass filter to the data to reduce noise and then process the following five dimensions for each accelerometer, gyroscope, or linear accelerometer sample: the filtered $x$-, $y$-, and $z$-values, the energy of those filtered values, and the energy of the unfiltered (raw) values, where the energy of $\{x, y, z\}$ is given by $\sqrt{x^2+y^2+z^2}$. As GRV samples are expressed as quaternions, for those we process only the four filtered values (since the Euclidean norm of a quaternion is always 1). In total, we process each gesture in 19 dimensions.

For each gesture, we extract the following ten statistical features in each dimension: \textit{minimum}, \textit{maximum}, \textit{mean}, \textit{median}, \textit{standard deviation}, \textit{variance}, \textit{inter-quartile range}, \textit{kurtosis}, \textit{skewness}, and \textit{peak count}. We also calculate the \textit{mean} and \textit{maximum velocities} along each axis, the \textit{displacement} along each axis, and the \textit{Euclidean displacement} from each of its accelerometer, gyroscope, and linear accelerometer vectors, adding another 30 features. Ultimately, we reduce each gesture to a feature vector containing 220 members.

We began with a larger set of features that had been used successfully by other authors in similar scenarios \cite{Acar2020,Cornelius2012,Mare2014,Ravi2005} and we pruned it down by using normalised Gini importances to reject the least informative features. The Gini importance of a feature is a measure of its effect to decrease the impurity of the model \cite{Breiman1984} and thus its positive impact on classification. We chose to include kurtosis and skewness due to an observation that in several dimensions there was one prominent peak whose shape or position correlated well per user. We chose to calculate velocity and displacement due to the variability in orientation of the watch in transit between the user and terminal.

\subsection{Classification}\label{sec:MethodsClassification}

We employ three supervised learning approaches, training separate models for authentication and intent recognition.

For our authentication model, we train a set of classifiers that are user-dependent and terminal-agnostic. In each, we take a given user's tap gestures as the positive class and other users' tap gestures as the negative class. As this is an authentication use-case, we ensure that the training data precedes the testing data by taking the tap gestures collected in users' first and second data collection sessions as training data (analogous to the enrolment phase, where the user template is created) and those collected in the third session as testing data (analogous to an authentication phase). For each user, we train multiple classifiers, each one excluding a different fixed terminal; we train the classifier on users' tap gestures performed on the other terminals and then test it on those performed on the excluded terminal, to ensure that the model is terminal-agnostic and generalised. With 16 users and 6 fixed terminals, this gives $16\times6 = 96$ separate classifiers.

Furthermore, for investigative purposes, we also train a \textit{terminal-specific} authentication model, in which each classifier is trained and tested on tap gestures performed on a single terminal. This model enables us to compare the effectiveness of dedicated, terminal-specific classifiers for systems that have standardised terminals, such as public transport systems.

For our intent recognition model, we train a set of tap gesture recognition classifiers that are not user-dependent. We take all users' tap gestures as the positive class and all users' non-tap gestures as the negative class---that is, we treat tap gestures performed on a terminal as intentional and other gestures as unintentional. As the two classes are highly imbalanced in this model (with the negative class being many times larger than the positive), we apply a stratified 10-fold cross-validation approach to preserve class proportionality in training and testing folds and to avoid bias towards the more populous class. For each user, we train classifiers using only the data of other users to ensure that they are user-agnostic.

We use random forest classifiers in each of our models. A random forest is an ensemble learning method that combines the efforts of multiple decision trees, each constructed from a randomly-selected, bootstrapped sample of training data, and outputs the modal class. Random forests have been shown to be efficient, able to estimate the importance of features, and robust against noise \cite{Caruana2006, Maxion2010}. To balance relevance with learning time, we include 100 trees in each forest \cite{Oshiro2012}. To reduce the impact of random generation on our results, and to avoid the deceptive practice of selecting results only from the most performant forest, we train and test each classifier ten times with different forest randomisation seeds and average the outcomes.

\subsection{Performance Metrics}

In each model, the \textit{true positives} is the number of times that the positive class (\textit{i.e.}, the legitimate user or intentional gesture) is correctly accepted; the \textit{true negatives} is the number of times that the negative class (\textit{i.e.}, the adversary or unintentional gesture) is correctly rejected; the \textit{false positives} is the number of times that the negative class is wrongly accepted; and the \textit{false negatives} is the number of times that the positive class is wrongly rejected.

To quantify the performance of our models and to compare our results with those of our closest related work \cite{Shrestha2016} using the same metrics, we calculate precision, recall, and F-measure.
Precision indicates \textit{security}, by measuring how well the model rejects the negative class, and recall indicates \textit{usability}, inasmuch as it measures how well the model avoids misclassifying the positive class and causing inconvenience to the user; F-measure is the harmonic mean of the precision and recall (equally weighted), offering a rough fusion of the two, which is ideal for our purposes as we want to consider a balance of both security and usability.

To quantify the performance of our models when used as an additional factor in an existing system, we want to measure the security benefit that we provide without incurring any cost to usability in the form of false negatives. To evaluate our models in this regard, we find for each model the optimum decision threshold that yields minimal false negatives and we measure the false acceptance rate (FAR) there. The FAR inversely indicates \textit{security}, by measuring the likelihood that the negative class will be wrongly accepted. The false rejection rate (FRR) inversely indicates \textit{usability}, by measuring the likelihood that the positive class will be wrongly rejected.

The decision threshold, $\theta$, is the score at which the classifier chooses to assign to a sample the positive class rather than the negative. To finely tune the classifier, we adjust $\theta$ to modify the trade-off between security and usability; a larger $\theta$ is more resilient to false positives and thus favours security, a smaller $\theta$ favours usability. To minimise the occurrence of false negatives, we optimise our models by selecting $\theta$ such that the FRR is less than 0.01\% and the corresponding FAR is as low as possible (an example is shown in Figure \ref{fig:ResultsDecisionThreshold}).

The FAR and FRR are antagonistic insofar as setting $\theta$ to favour one will disfavour the other. The crossover point is called the equal error rate (EER) and is a measure of system performance when consideration is balanced evenly between security and usability. To measure the impact of optimisation on a model (\textit{i.e.}, how much of the potential security gains are sacrificed to minimise the impact on usability), we compare the FAR when optimised with the EER (and therefore the FAR) when not.

\begin{figure*}[t!]
	\centering
	\begin{subfigure}[t]{0.44\textwidth}
		\centering
		\includegraphics[height=6.4cm]{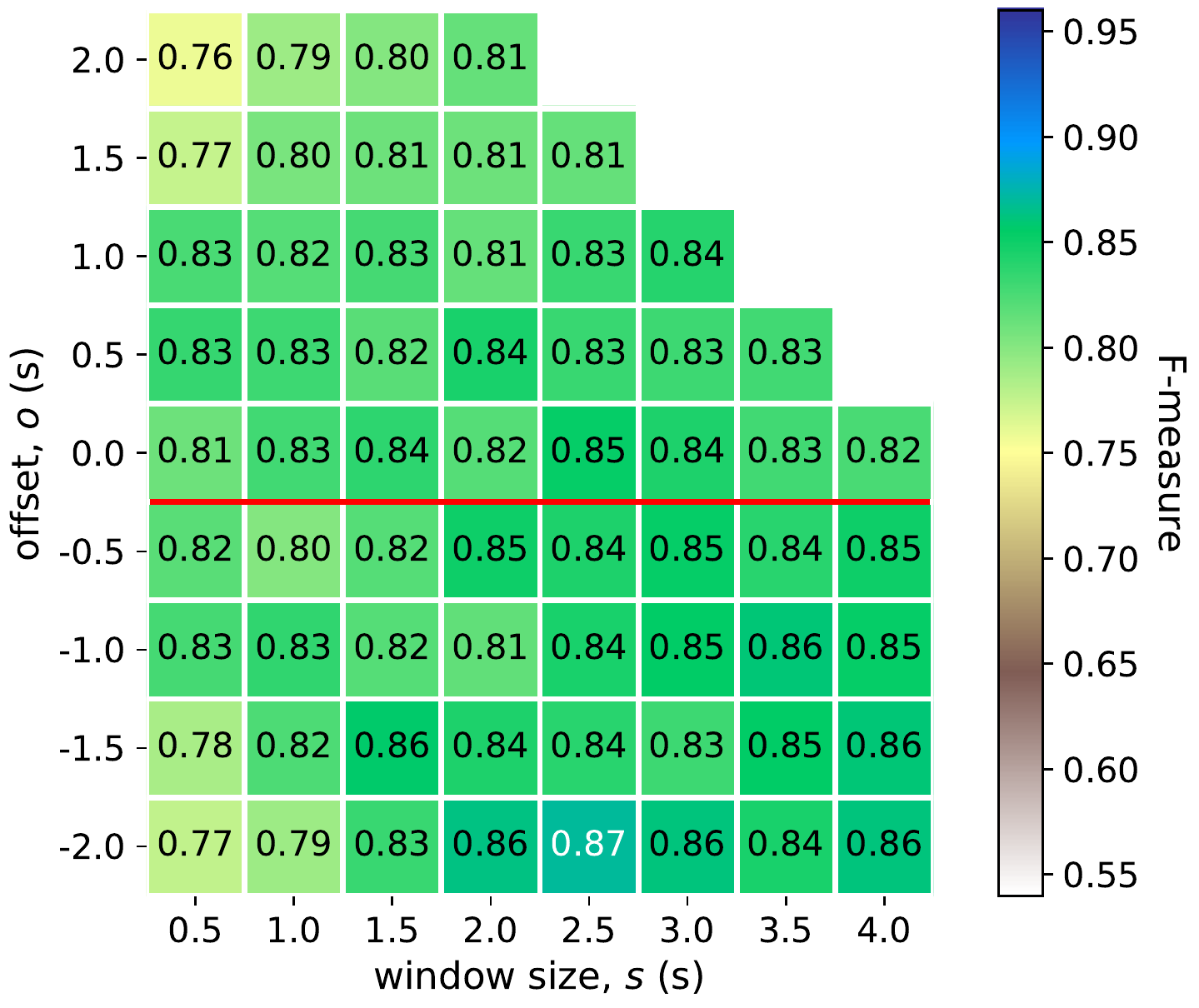}
		\caption{authentication model}
		\label{fig:ResultsF1Auth}
	\end{subfigure}
	~
	\begin{subfigure}[t]{0.44\textwidth}
		\centering
		\includegraphics[height=6.4cm]{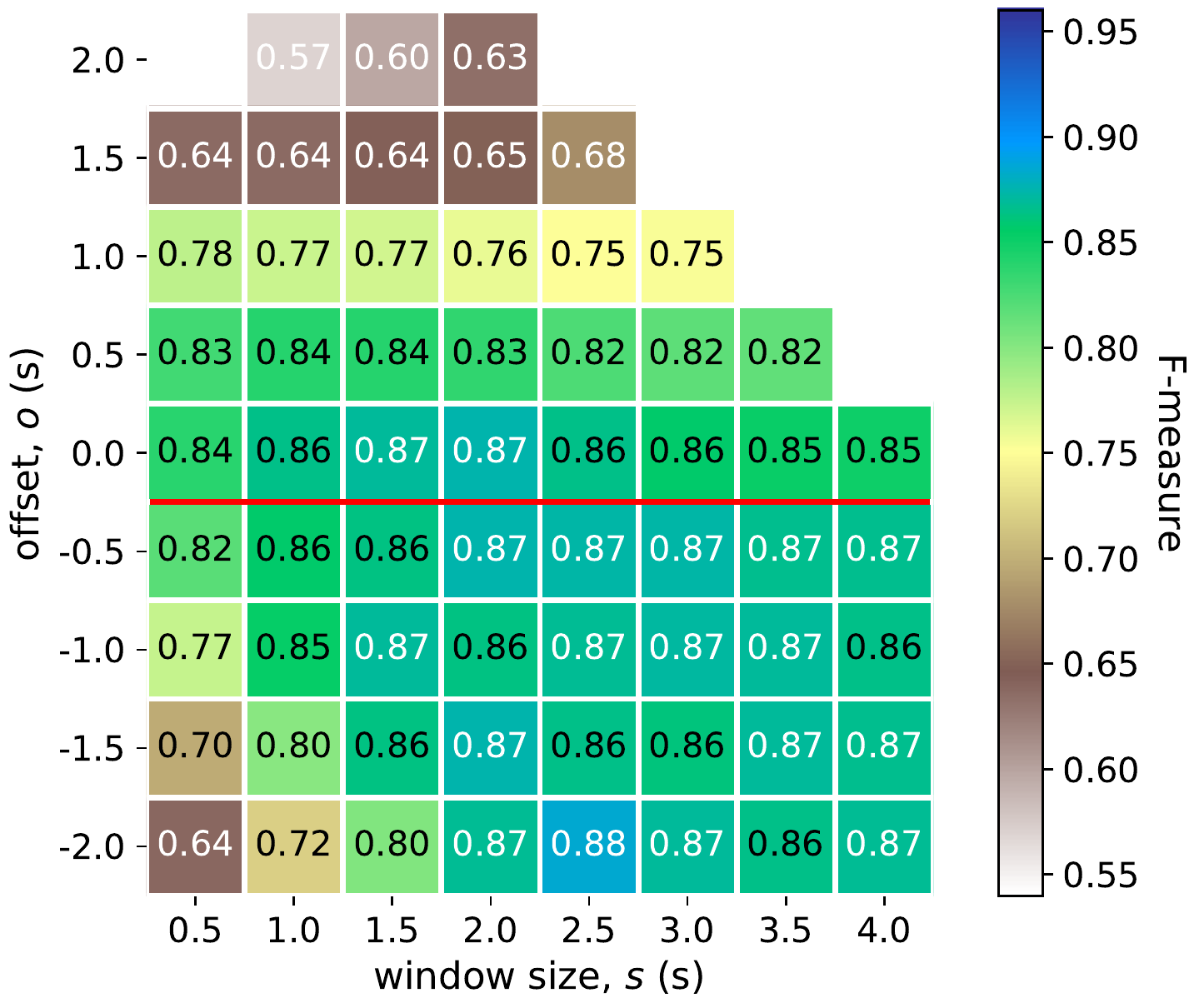}
		\caption{intent recognition model}
		\label{fig:ResultsF1Recog}
	\end{subfigure}
	\caption{Average F-measure scores for our authentication and intent recognition models by window size and offset. Tap gestures that end at or before the NFC contact point, which are therefore compatible with in-store usage, are above the red line.}
	\label{fig:ResultsF1}
\end{figure*}

\section{Results}\label{sec:Results}

\subsection{Anatomy of a Tap Gesture}

We observe that a tap gesture can be demarcated into the following three phases: \textit{reaching}, \textit{alignment}, and \textit{withdrawal}. In the reaching phase, the user extends his arm to move the smartwatch towards the terminal. Once the watch is touching or very close to touching the terminal, the user stops reaching and enters the alignment phase. In the alignment phase, the user aligns the watch face with the terminal and tentatively moves it around to find an NFC connection, owing to the short-ranged nature of NFC technology. Once a connection is established and the payment is approved, the terminal notifies the user with a sound or message and the user withdraws. We find that the alignment phase for a typical tap gesture in our study begins 0.5 to 1.5 seconds before the NFC contact point, depending on how quickly the NFC connection is established, and ends 0 to 1 seconds after, depending on how quickly the user reacts to the notification.

\subsection{Optimum Window Parameters}\label{sec:ResultsOptimumWindowParameters}

Figure \ref{fig:ResultsF1} shows the F-measure scores for our authentication and intent recognition models by window size and offset. Each score in Figure \ref{fig:ResultsF1Auth} is the average of scores from $16\times6\times10 = 960$ classifiers and in Figure \ref{fig:ResultsF1Recog}, from $16\times10 = 160$ (see Section \ref{sec:MethodsClassification} for details).

\textbf{In-store Usage.} For real-time usage, we focus only on offsets $o\ge0$ (above the red line), where the tap gestures end at or before the NFC contact point, so the system can use the result in deciding whether to approve the payment.

For authentication, we see in Figure \ref{fig:ResultsF1Auth} that our model achieves an average F-measure score of 0.83 with very little deviation across window sizes with offsets $o\le1$; our best score is 0.85 at $\{s=2.5, o=0\}$. We find that even 0.5 seconds of wrist motion data is sufficient to authenticate the user with a tap gesture.

For intent recognition, we find in Figure \ref{fig:ResultsF1Recog} that offset is the determining factor and that smaller offsets yield stronger results; the model achieves its best in-store results when $o=0$, with an average F-measure score of 0.86.

Considering Figures \ref{fig:ResultsF1Auth} and \ref{fig:ResultsF1Recog} together, we find that the optimum window parameters for in-store usage, favouring authentication, are $\{s=2.5, o=0\}$ (\textit{i.e.}, a window of 2.5 seconds of sensor data taken immediately before the NFC contact point). Using a single tap gesture so windowed, our models can authenticate the user with an average F-measure score of 0.85 (precision 0.92, recall 0.84, EER 0.10) and recognise intent-to-pay with an average F-measure score of 0.86 (precision 0.93, recall 0.82, EER 0.04).

\textbf{Retrospective Fraud Detection.} For historic analysis, we consider all results in the heatmaps. We see a slight improvement in F-measure scores for windows containing sensor data collected after the NFC contact point, especially where the tap gesture spans it (\textit{i.e.}, where $o<0$ and $s>|o|$); this suggests that a user's withdrawal is at least as distinctive as his movement towards the terminal. The optimum window parameters here are $\{s=2.5, o=-2\}$.

For authentication, our results suggest that the withdrawal phase is more distinctive between users than the alignment phase, as windows that contain more data from that phase tend to yield stronger results, although the differences are too small to be conclusive. We can see that the EERs in Figure \ref{fig:ResultsEERAuth} corroborate this; they also suggest that larger window sizes produce a better balance of security and usability. The results begin to decline at the top- and bottom-left corners of Figure \ref{fig:ResultsF1Auth}; these windows have the highest likelihood of containing data that is irrelevant to the tap gesture, data that is collected from random movements before or after the tap gesture, respectively, and is therefore harder to classify.

For intent recognition, we see that the alignment phase of the gesture is the most distinctive between gesture types. We see a strong correlation between the preponderance of alignment phase data in a window and the strength of its results. This is most evident in the inverse, as we see that the fewer alignment phase samples a window has, the weaker its results: at the top of Figure \ref{fig:ResultsF1Recog}, the larger the positive offset, the fewer alignment phase samples it is likely to contain, and at the bottom-left, the larger the negative offset and the smaller the window, the fewer alignment phase samples it is likely to contain. The constricted movement as the watch moves in conformity to the surface of the terminal and the manner in which the user reacts to finding the NFC connection are peculiar to the tap gesture and so act to distinguish it from other gestures. Strong results are given by those windows that span all three phases, in particular $\{2\le s\le3, o=-0.5\}$.

\subsection{Feature Informativeness}\label{sec:ResultsFeatureInformativeness}

To see which features are most informative to our models, we sum the top five features, sorted by Gini importance, of each classifier. Table \ref{tab:ResultsModalFeatures} shows the modal top-five features summed over classifiers with optimum window parameters $\{s=2.5, o=0\}$ and across all windows. (Note that, \textit{w.r.t.} the counts, there are six times more classifiers for authentication.)

For authentication, we see in Table \ref{tab:ResultsModalFeaturesAuth} that features derived from the $y$-axis of the gyroscope are common among the most informative; this suggests that the forward roll of the wrist is a key discriminator between users. The extremes in acceleration along the $x$-axis, representing the rapidity of the extension and withdrawal of the arm, is also shown to be important.

For intent recognition, we see in Table \ref{tab:ResultsModalFeaturesRecog} that the number of peaks in the magnitude of linear accelerometer samples is of particular importance, far exceeding any other in the count across all windows. This feature represents the frequency with which the watch starts and stops moving during the tap gesture and is prominent here likely owing to the significance of the alignment phase data to this model and the abrupt movements performed during that phase. It is notable that there are no GRV-derived features among the commonest (indeed, we also tallied the top-twenty features and saw no GRV-derived features present there either). Together, these findings suggest that the distinctiveness of the alignment phase does not come from the orientation of the watch face, but from the \textit{changes} in orientation detected across sensors. Features that are derived from the $x$-value of the gyroscope data, which measures the tilt of the wrist from side to side (see Figure \ref{fig:ExperimentSensors}), likely express this most profoundly (in particular, \texttt{Gyr-x-velomean} gives a running approximation of the sideways orientation of the device). We also see that features derived from the $z$-value of the accelerometer data are frequently among the most important in distinguishing between gestures; this is likely to be because sustained movement in the direction of the watch face is peculiar to the tap gesture.

\subsection{Sensor Selection}

\begin{figure*}[t!]
	\textcolor{white}{\\.\\.\\.}
	\centering
	\begin{subfigure}[t]{0.44\textwidth}
		\centering
		\includegraphics[height=6.4cm]{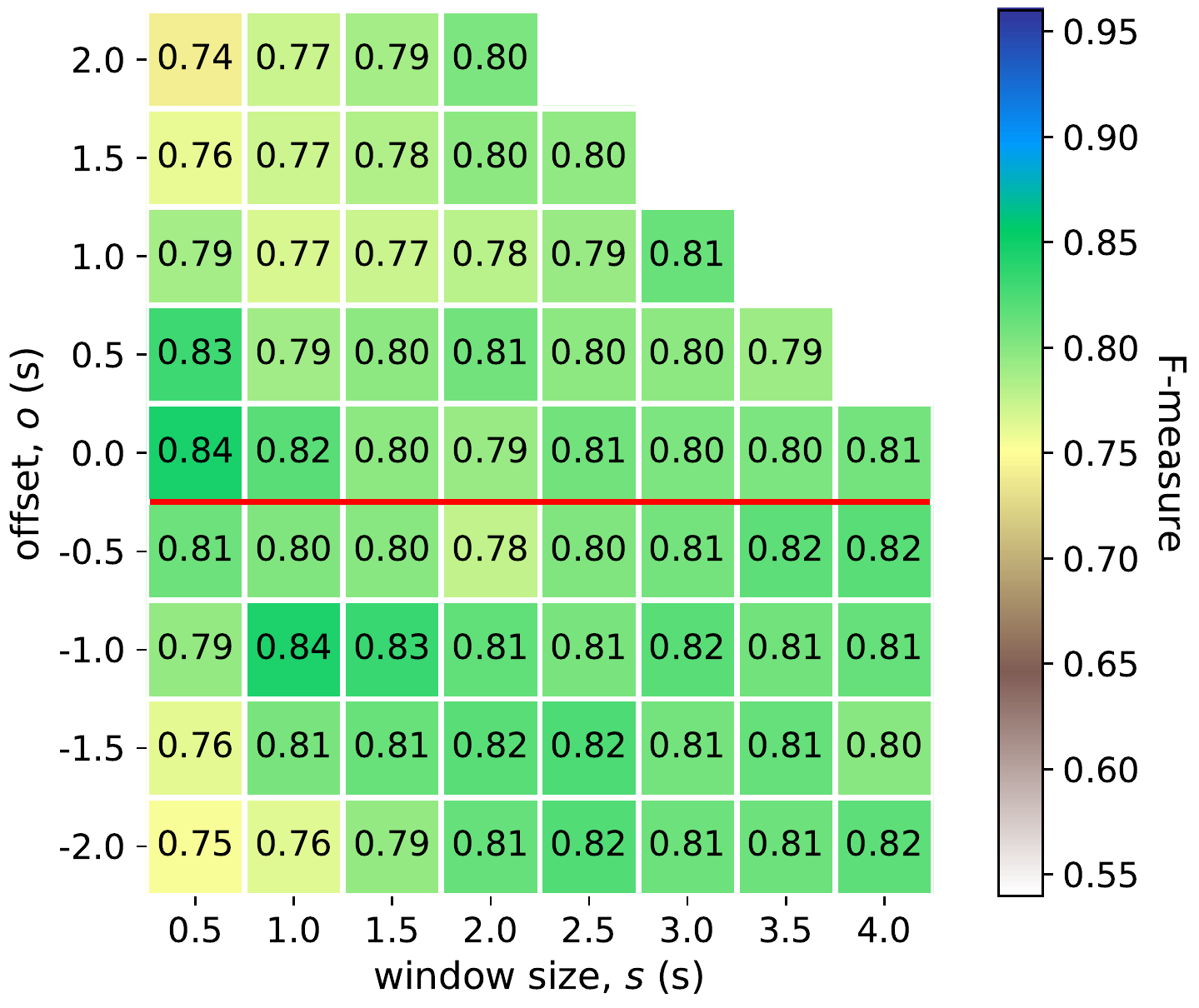}
		\caption{authentication model; accelerometer \& gyroscope\textcolor{white}{\\.}}
		\label{fig:ResultsSensorsAccGyrAuth}
	\end{subfigure}
	~
	\begin{subfigure}[t]{0.44\textwidth}
		\centering
		\includegraphics[height=6.4cm]{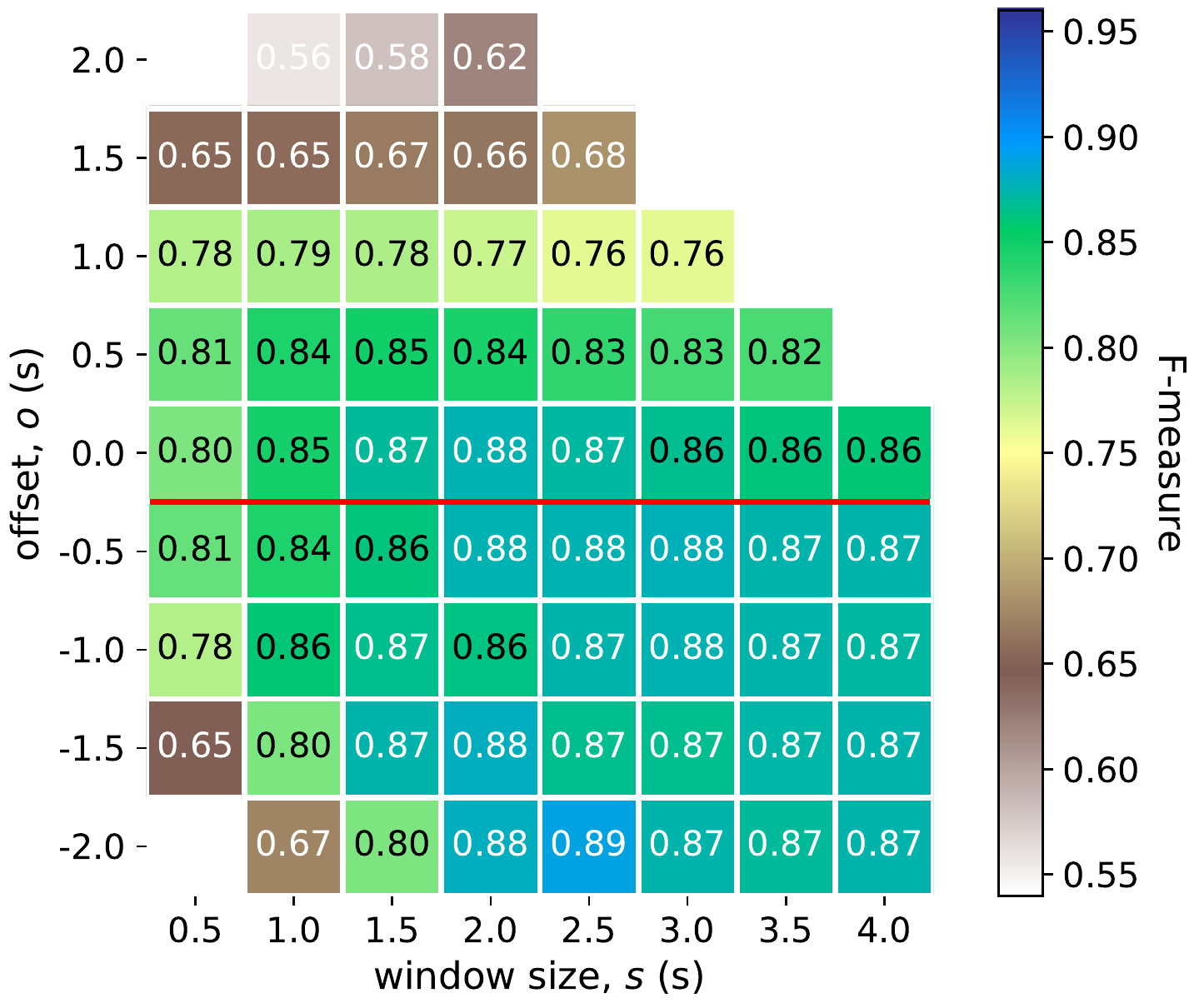}
		\caption{intent recognition model; accelerometer \& gyroscope\textcolor{white}{\\.}}
		\label{fig:ResultsSensorsAccGyrRecog}
	\end{subfigure}
	~
	\begin{subfigure}[t]{0.44\textwidth}
		\centering
		\includegraphics[height=6.4cm]{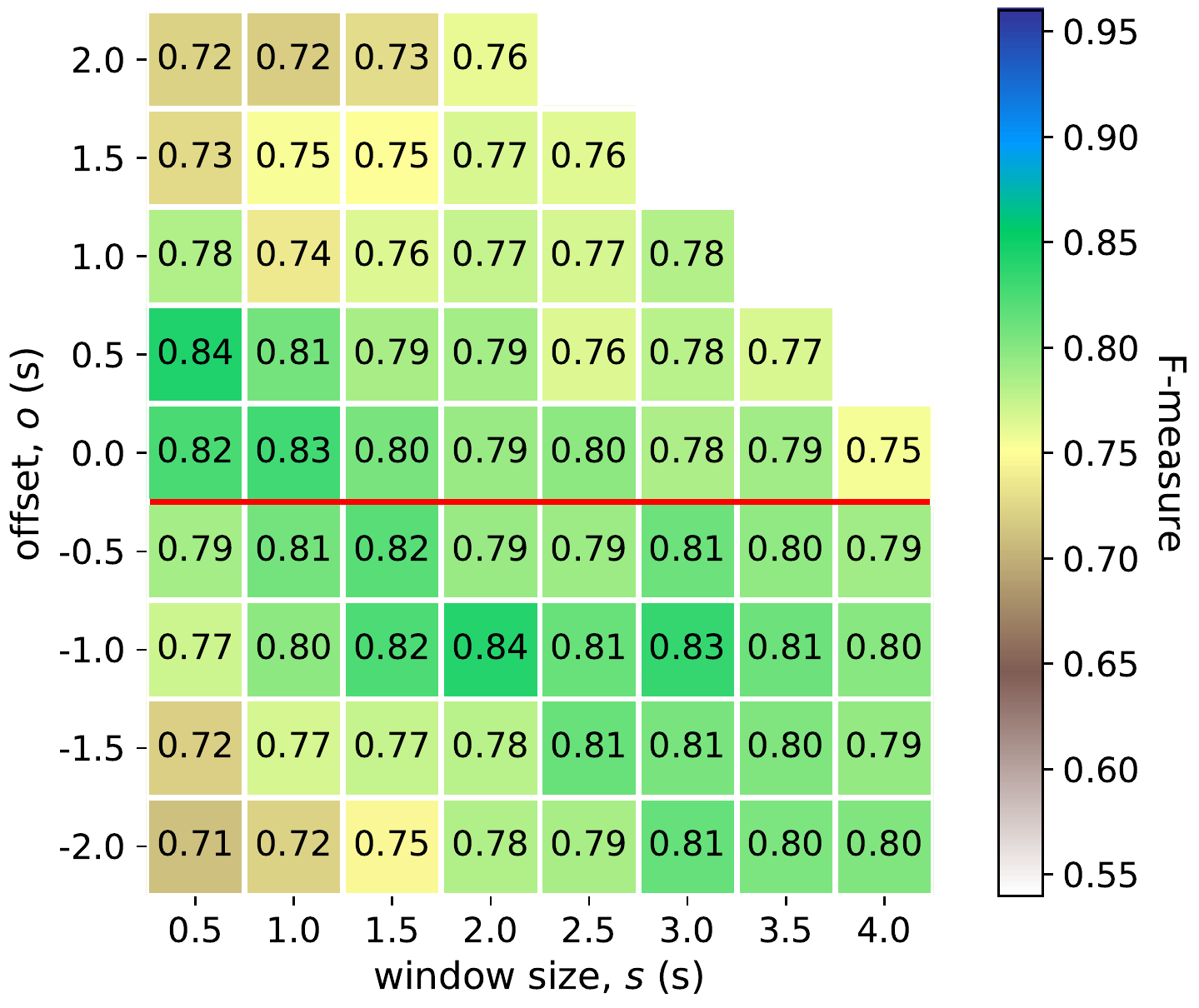}
		\caption{authentication model; accelerometer}
		\label{fig:ResultsSensorsAccAuth}
	\end{subfigure}
	~
	\begin{subfigure}[t]{0.44\textwidth}
		\centering
		\includegraphics[height=6.4cm]{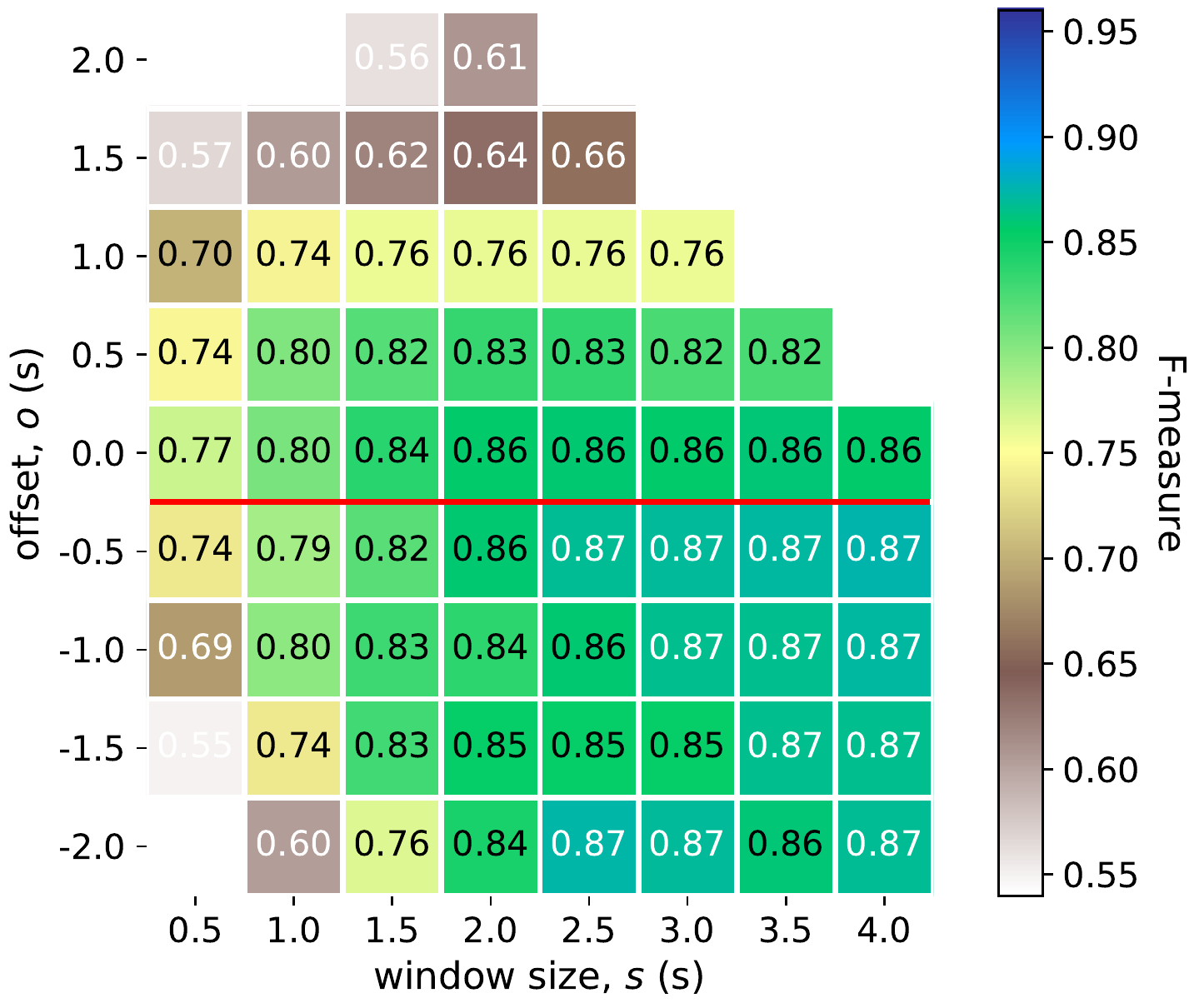}
		\caption{intent recognition model; accelerometer}
		\label{fig:ResultsSensorsAccRecog}
	\end{subfigure}
	\caption{Average F-measure scores for our authentication and intent recognition models by window size, offset, and sensors. Tap gestures that end at or before the NFC contact point, which are therefore compatible with in-store usage, are above the red line.\textcolor{white}{\\.\\.\\.}}
	\label{fig:ResultsSensors}
\end{figure*}

We collected wrist motion data from all four of the inertial sensors available on our smartwatch. Some devices are more limited in their offering---the accelerometer is the commonest sensor, as it is the smallest and cheapest, followed by the gyroscope. To gauge the feasibility of our approach on devices with fewer sensors, we trained a set of sensor-specific models in which each classifier is trained and tested on data from a subset of sensors. Figure \ref{fig:ResultsSensors} shows the F-measure scores for models using data from (i) the accelerometer and gyroscope and (ii) only the accelerometer (\textit{cf.} Figure \ref{fig:ResultsF1}, which uses all four).

\begin{table}[t!]
	\centering
	\small
	\begin{subtable}{0.48\textwidth}
		\centering
		\begin{tabular}{c|c|c|c}
        	\toprule
        	\multicolumn{2}{c|}{\textbf{$s=2.5, o=0$}} & \multicolumn{2}{c}{\textbf{All Windows}} \\
        	\midrule
        	\textbf{Feature} & \textbf{Count} & \textbf{Feature} & \textbf{Count} \\
        	\midrule
        	\texttt{Acc-x-min} & 218 & \texttt{Acc-x-min} & 9261 \\
        	\texttt{Gyr-y-velomean} & 207 & \texttt{Acc-x-max} & 9225 \\
       		\texttt{Gyr-y-mean} & 178 & \texttt{Gyr-y-mean} & 8129 \\
        	\texttt{Gyr-y-disp} & 169 & \texttt{Acc-x-velomean} & 7513 \\
        	\texttt{Gyr-y-max} & 146 & \texttt{Gyr-y-velomean} & 6459 \\
        	\texttt{Gyr-y-med} & 143 & \texttt{GRV-x-min} & 6304 \\
        	\texttt{Gyr-y-velomax} & 125 & \texttt{Gyr-y-med} & 6234 \\
        	\texttt{Acc-x-max} & 121 & \texttt{GRV-x-mean} & 6085 \\
        	\texttt{Gyr-z-max} & 120 & \texttt{Gyr-y-velomax} & 6018 \\
        	\texttt{GRV-x-min} & 116 & \texttt{Gyr-y-disp} & 5627 \\
        	\bottomrule
		\end{tabular}
		\caption{authentication model\textcolor{white}{\\.}}
		\label{tab:ResultsModalFeaturesAuth}
	\end{subtable}
	~
	\begin{subtable}{0.48\textwidth}
		\centering
		\begin{tabular}{c|c|c|c}
        	\toprule
        	\multicolumn{2}{c|}{\textbf{$s=2.5, o=0$}} & \multicolumn{2}{c}{\textbf{All Windows}} \\
        	\midrule
        	\textbf{Feature} & \textbf{Count} & \textbf{Feature} & \textbf{Count} \\
       		\midrule
        	\texttt{Acc-z-iqr} & 79 & \texttt{LAc-unf-pkcount} & 3397 \\
        	\texttt{Acc-z-kurt} & 65 & \texttt{Gyr-x-mean} & 1985 \\
        	\texttt{Gyr-x-mean} & 57 & \texttt{Acc-z-var} & 1789 \\
        	\texttt{LAc-unf-pkcount} & 55 & \texttt{Acc-z-stdev} & 1706 \\
        	\texttt{Acc-disptotal} & 44 & \texttt{Acc-z-iqr} & 1696 \\
        	\texttt{Acc-unf-iqr} & 43 & \texttt{Acc-z-med} & 1316 \\
        	\texttt{Gyr-unf-min} & 43 & \texttt{Acc-unf-iqr} & 1118 \\
        	\texttt{Acc-z-var} & 33 & \texttt{Gyr-x-velomean} & 1011 \\
        	\texttt{Acc-z-stdev} & 20 & \texttt{Gyr-unf-min} & 971 \\
        	\texttt{Acc-x-pkcount} & 11 & \texttt{Acc-disptotal} & 907 \\
        	\bottomrule
		\end{tabular}
		\caption{intent recognition model\textcolor{white}{\\.}}
		\label{tab:ResultsModalFeaturesRecog}
	\end{subtable}
	~
	\begin{subtable}{0.49\textwidth}
		\centering
		\begin{tabular}{c|c|c|c}
        	\toprule
        	\multicolumn{2}{c|}{\textbf{$s=2.5, o=0$}} & \multicolumn{2}{c}{\textbf{All Windows}} \\
        	\midrule
        	\textbf{Feature} & \textbf{Count} & \textbf{Feature} & \textbf{Count} \\
        	\midrule
        	\texttt{Acc-z-kurt} & 73 & \texttt{Acc-z-var} & 2109 \\
        	\texttt{Gyr-unf-min} & 73 & \texttt{Acc-unf-iqr} & 2049 \\
        	\texttt{Acc-z-iqr} & 70 & \texttt{Gyr-x-mean} & 1841 \\
        	\texttt{Acc-unf-iqr} & 65 & \texttt{Acc-z-stdev} & 1716 \\
        	\texttt{Gyr-x-mean} & 52 & \texttt{Acc-z-iqr} & 1485 \\
        	\texttt{Acc-z-var} & 47 & \texttt{Acc-unf-pkcount} & 1412 \\
        	\texttt{Acc-disptotal} & 35 & \texttt{Acc-z-med} & 1368 \\
        	\texttt{Acc-z-stdev} & 30 & \texttt{Gyr-unf-min} & 1239 \\
        	\texttt{Acc-x-pkcount} & 3 & \texttt{Gyr-x-disp} & 1117 \\
        	\texttt{Gyr-x-disp} & 2 & \texttt{Gyr-x-velomean} & 1043 \\
        	\bottomrule
		\end{tabular}
		\caption{intent recognition model; accelerometer \& gyroscope}
		\label{tab:ResultsModalFeaturesRecogAccGyr}
	\end{subtable}
	\caption{Modal top-five features by Gini importance summed over all classifiers in optimum window $\{s=2.5, o=0\}$ and across all windows for our authentication and intent recognition models (for the latter, once trained and tested on all sensor data and once on a subset). Features are given in the format sensor-axis-statistic; \texttt{unf} is the magnitude of the unfiltered $\{x, y, z\}$ values and \texttt{disptotal} is the Euclidean displacement.}
	\label{tab:ResultsModalFeatures}
\end{table}

For authentication, we see that there is a monotonic improvement in results the more sensors are included, with few exceptions.

For intent recognition, we see comparable results across all windows in Figures \ref{fig:ResultsF1Recog} and \ref{fig:ResultsSensorsAccRecog}, but improved results up to 0.89 in Figure \ref{fig:ResultsSensorsAccGyrRecog}; this suggests that the inclusion of the linear accelerometer and GRV is unnecessary and pollutes the classifiers. Table \ref{tab:ResultsModalFeaturesRecogAccGyr} shows the modal top-five features for the intent recognition model with these sensors omitted; compared with Table \ref{tab:ResultsModalFeaturesRecog}, we see that the frequency of starts and stops remains important, with \texttt{Acc-unf-pkcount} rising in prominence in the absence of \texttt{LAc-unf-pkcount} (although not to the same extent, per the count); the rest of the list is largely unchanged.

\begin{table*}[h!]
	\centering
	\small
	\begin{subtable}{0.44\textwidth}
		\centering
		\begin{tabular}{c|c|c|c|c}
        	\toprule
        	 & \multicolumn{4}{c}{\textbf{F-measure}} \\
        	\textbf{Terminal} & $s=1.5$ & $s=2$ & $s=2.5$ & $s=3$ \\
       		\midrule
        	1 & 0.79 & 0.80 & 0.81 & 0.80 \\
        	2 & 0.77 & 0.77 & 0.77 & 0.80 \\
        	3 & 0.86 & 0.85 & 0.86 & 0.86 \\
        	4 & 0.79 & 0.80 & 0.81 & 0.81 \\
        	5 & 0.84 & 0.88 & 0.90 & 0.88 \\
        	6 & 0.81 & 0.84 & 0.86 & 0.85 \\
        	F & 0.86 & 0.86 & 0.86 & 0.87 \\
       		\midrule
        	agnostic & 0.84 & 0.82 & 0.85 & 0.84 \\
        	\bottomrule
		\end{tabular}
		\caption{terminal-specific authentication model}
		\label{tab:ResultsTerminalSpecificAuth}
	\end{subtable}
	~
	\begin{subtable}{0.44\textwidth}
		\centering
		\begin{tabular}{c|c|c|c|c}
        	\toprule
        	 & \multicolumn{4}{c}{\textbf{F-measure}} \\
        	\textbf{Terminal} & $s=1.5$ & $s=2$ & $s=2.5$ & $s=3$ \\
        	\midrule
        	1 & 0.88 & 0.87 & 0.86 & 0.86 \\
        	2 & 0.67 & 0.72 & 0.68 & 0.68 \\
        	3 & 0.65 & 0.65 & 0.68 & 0.67 \\
        	4 & 0.81 & 0.80 & 0.80 & 0.79 \\
        	5 & 0.81 & 0.81 & 0.80 & 0.79 \\
        	6 & 0.69 & 0.65 & 0.73 & 0.67 \\
        	F & 0.66 & 0.73 & 0.81 & 0.90 \\
        	\midrule
        	agnostic & 0.87 & 0.87 & 0.86 & 0.86 \\
        	\bottomrule
		\end{tabular}
		\caption{intent recognition model}
		\label{tab:ResultsTerminalSpecificRecog}
	\end{subtable}
	\caption{Average F-measure scores for our terminal-specific authentication and intent recognition models with optimum window parameters ($o=0$ in each case) for in-store usage by terminal. Terminal-agnostic results are included for comparison.\textcolor{white}{\\.\\.}}
	\label{tab:ResultsTerminalSpecific}
\end{table*}

\begin{figure*}[t!]
	\textcolor{white}{\\.\\.\\.}
	\centering
	\begin{subfigure}[t]{0.44\textwidth}
		\centering
		\includegraphics[height=6.4cm]{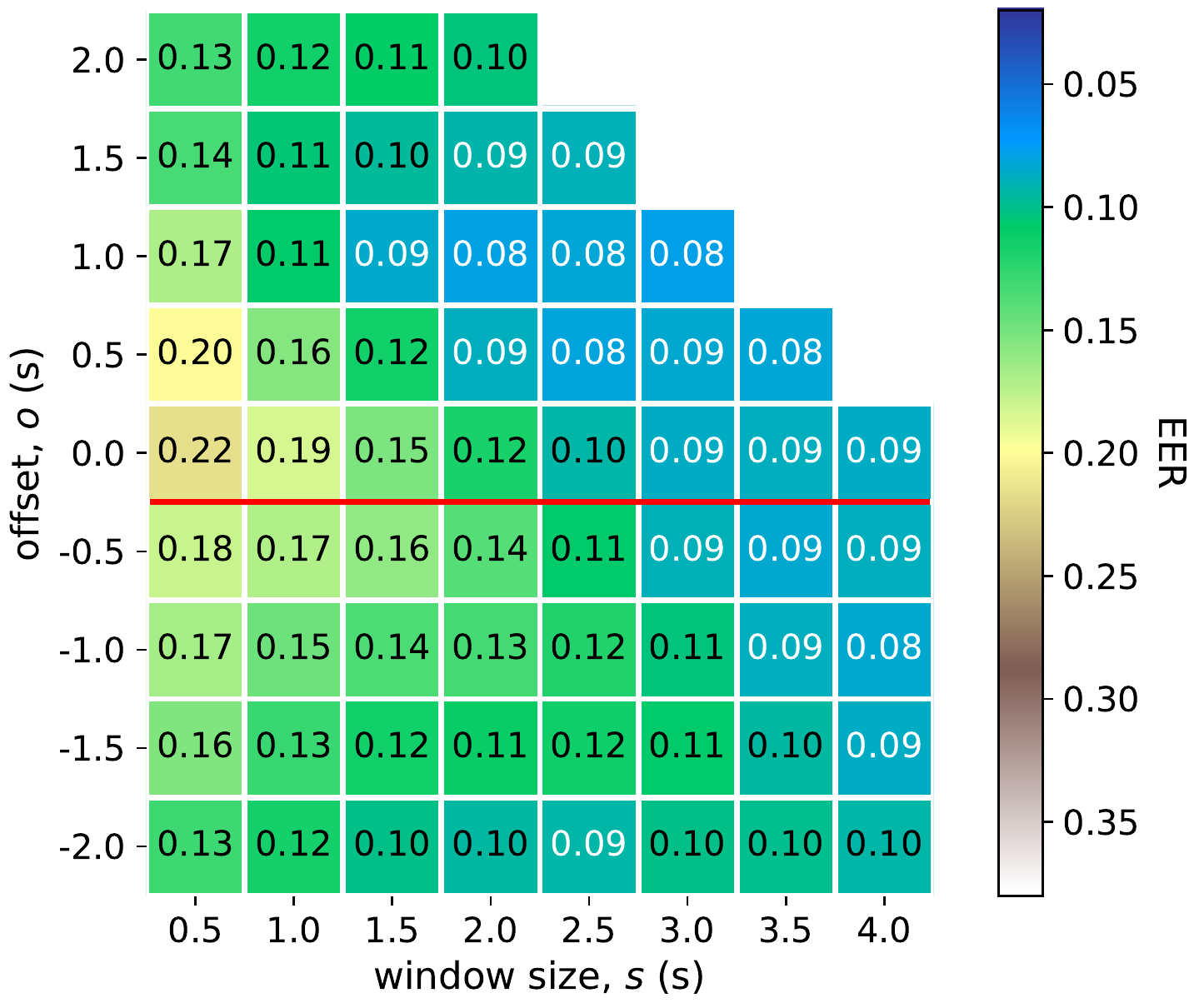}
		\caption{authentication model; EERs\textcolor{white}{\\.}}
		\label{fig:ResultsEERAuth}
	\end{subfigure}
	~
	\begin{subfigure}[t]{0.44\textwidth}
		\centering
		\includegraphics[height=6.4cm]{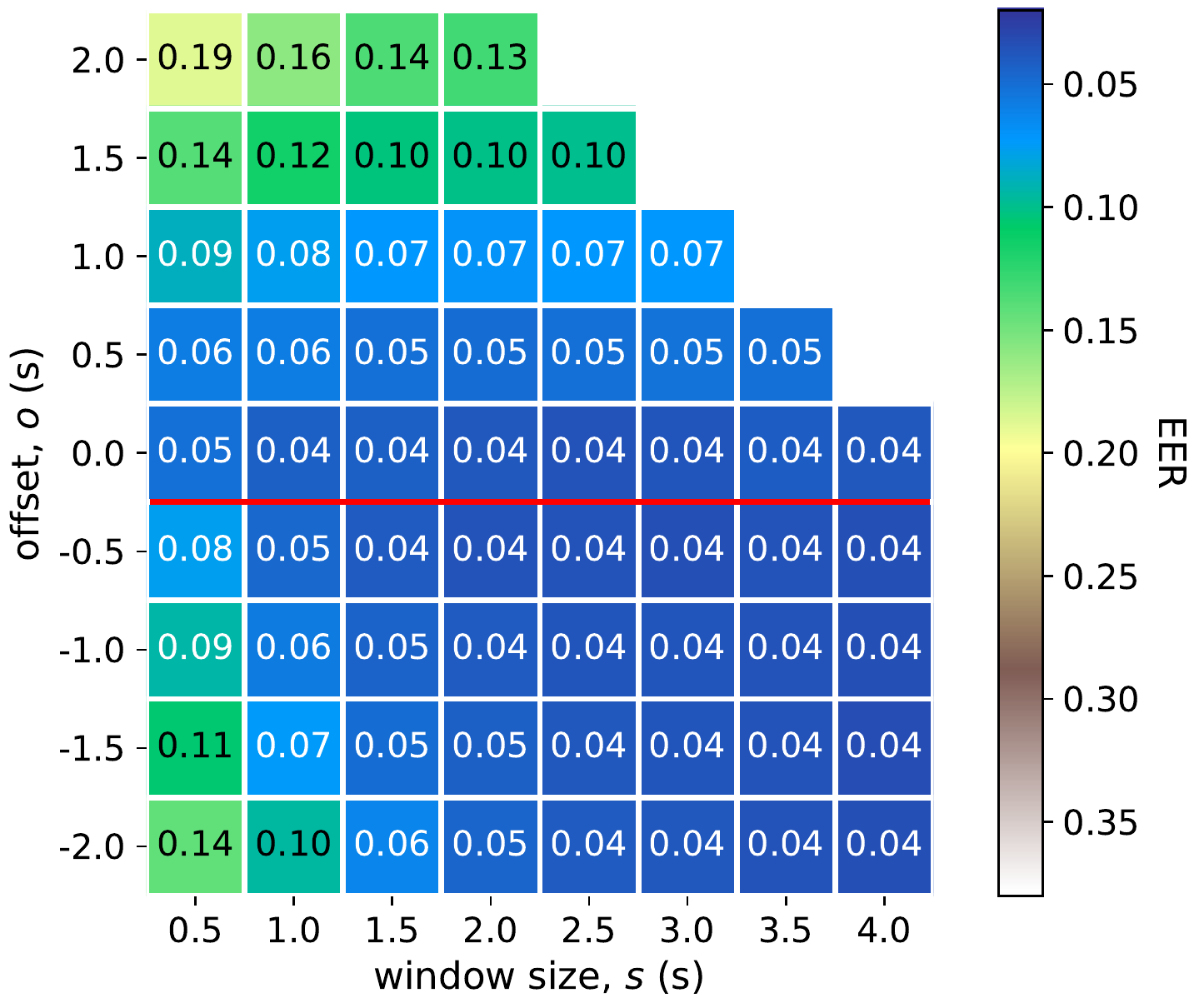}
		\caption{intent recognition model; EERs\textcolor{white}{\\.}}
		\label{fig:ResultsEERRecog}
	\end{subfigure}
	~
	\begin{subfigure}[t]{0.44\textwidth}
		\centering
		\includegraphics[height=6.4cm]{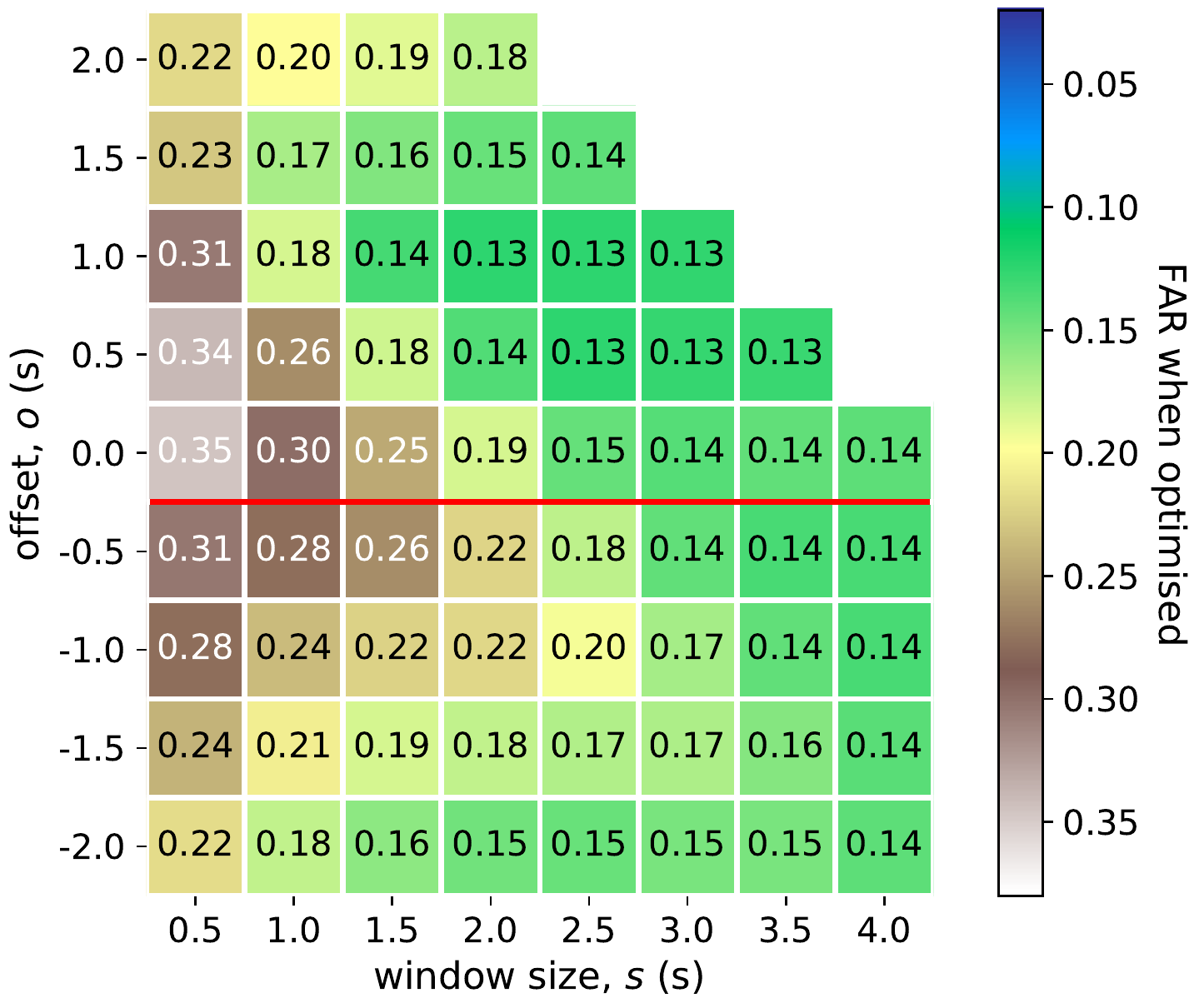}
		\caption{authentication model; FARs when optimised\textcolor{white}{\\.}}
		\label{fig:ResultsFARAuth}
	\end{subfigure}
	~
	\begin{subfigure}[t]{0.44\textwidth}
		\centering
		\includegraphics[height=6.4cm]{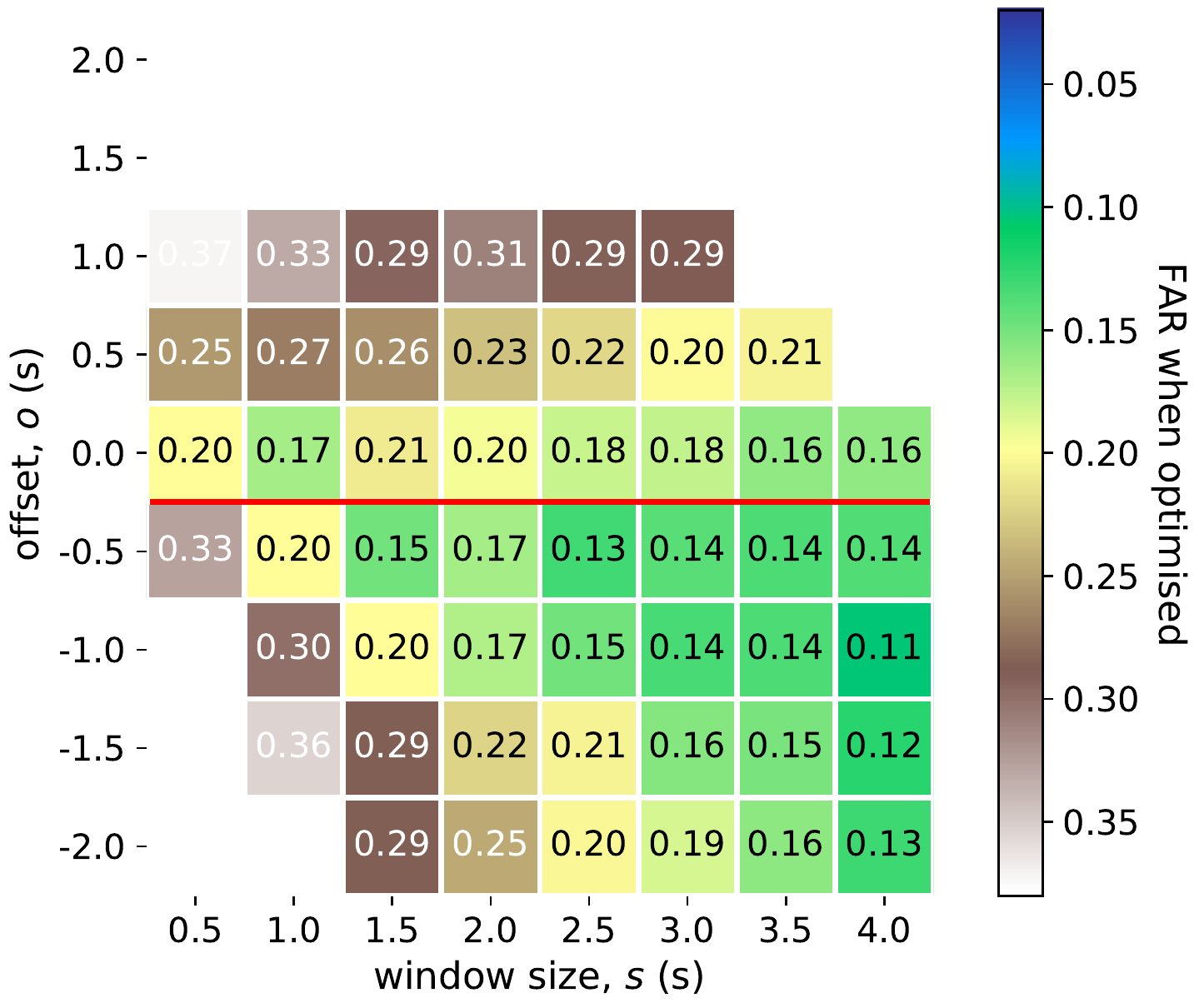}
		\caption{intent recognition model; FARs when optimised\textcolor{white}{\\.}}
		\label{fig:ResultsFARRecog}
	\end{subfigure}
	~
	\begin{subfigure}[t]{0.44\textwidth}
		\centering
		\includegraphics[height=6.4cm]{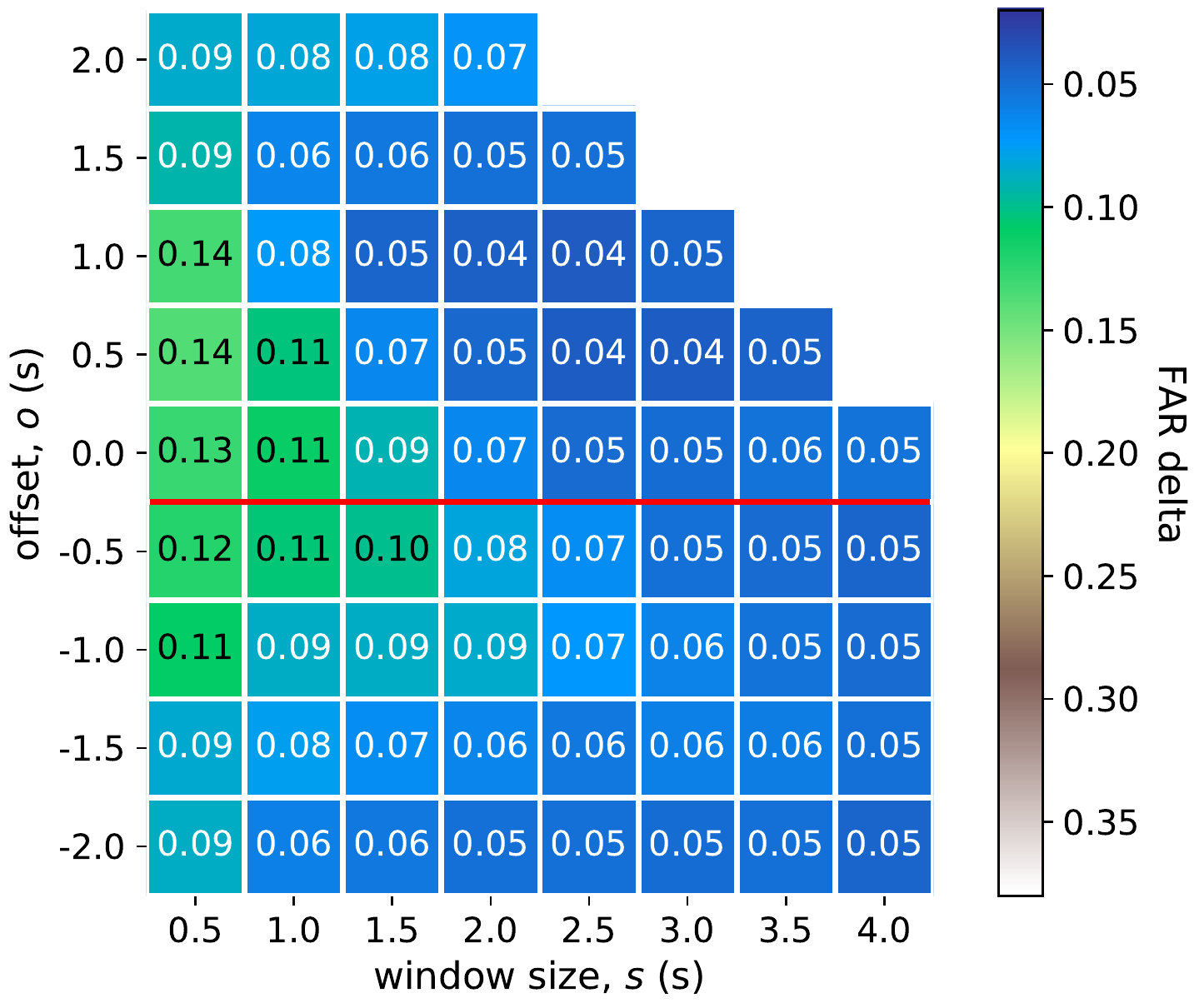}
		\caption{authentication model; FAR deltas}
		\label{fig:ResultsFARdeltaAuth}
	\end{subfigure}
	~
	\begin{subfigure}[t]{0.44\textwidth}
		\centering
		\includegraphics[height=6.4cm]{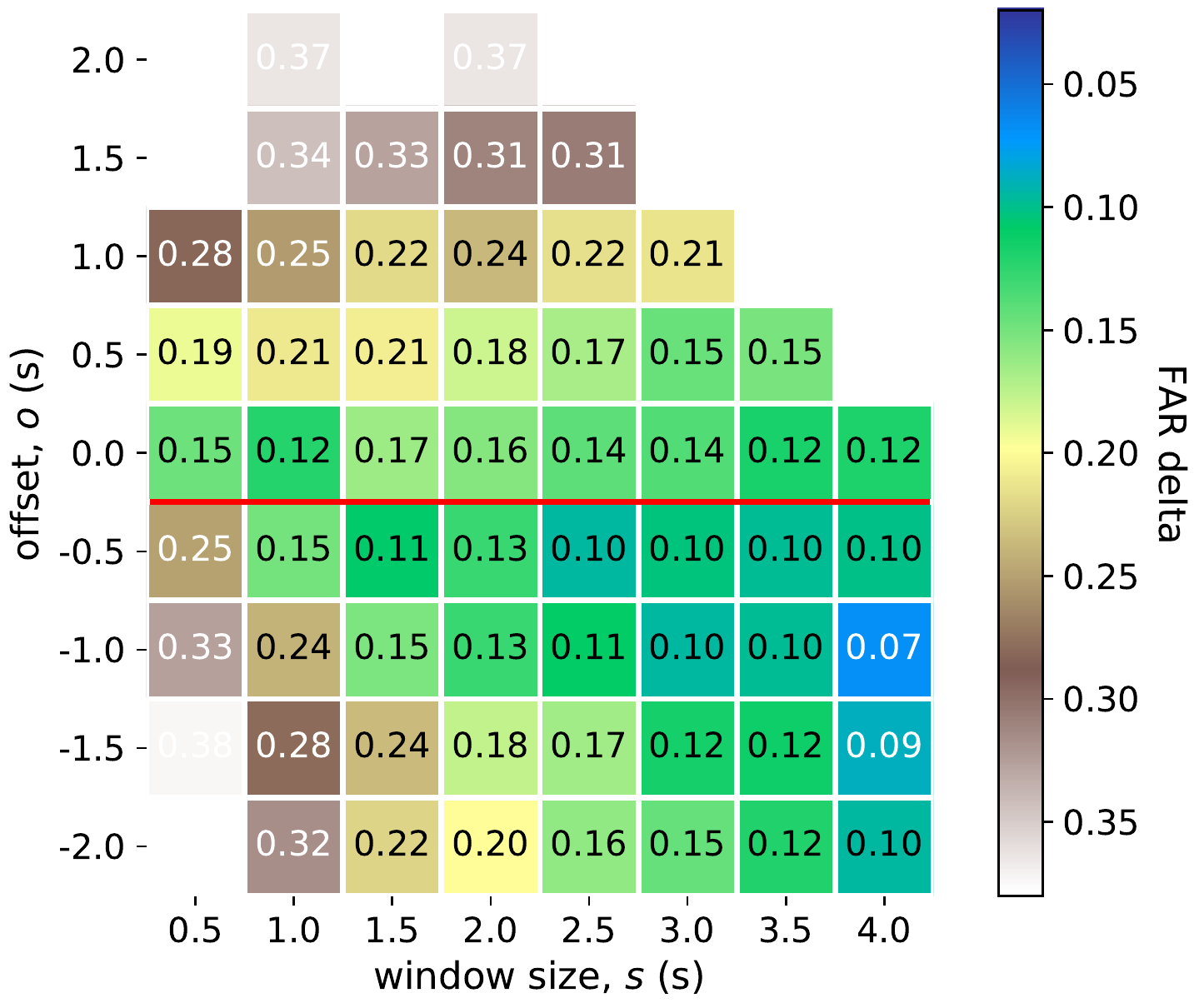}
		\caption{intent recognition model; FAR deltas}
		\label{fig:ResultsFARdeltaRecog}
	\end{subfigure}
	\caption{Average EERs, FARs when optimised for minimal false negatives, and the differences between these values (indicating the cost of optimisation) for our authentication and intent recognition models by window size and offset. Tap gestures that end at or before the NFC contact point, which are therefore compatible with in-store usage, are above the red line.\textcolor{white}{\\.\\.}}
	\label{fig:ResultsEER}
\end{figure*}

\subsection{Terminal Positions}

Table \ref{tab:ResultsTerminalSpecific} shows the F-measure scores for our terminal-specific authentication and intent recognition models, in which each classifier is trained and tested on tap gestures performed on a single terminal, for the four optimum windows identified above for in-store usage. We also include the terminal-agnostic results from our core authentication model, trained on tap gestures performed on all terminals other than the one under test, for comparison.

For authentication, we see in Table \ref{tab:ResultsTerminalSpecificAuth} that, per terminal, a similar trend is presented across the different window sizes, suggesting that window size is a more important factor than terminal position. We see that the results for Terminal 3 and the freestyle terminal are consistently better than those for the terminal-agnostic model. Terminal 3 protrudes towards the user and elicited a change in pose from users in the study as they interacted with it, resulting in a smoother, more comfortable tap gesture; the freestyle terminal accommodated this as well. The strength of results for these two terminals suggest that the reaching phase is more significant in the terminal-specific classifiers, perhaps because it is less constrained than the alignment phase and so offers the opportunity for user-distinctive traits to present. This appears to be less pronounced in the terminal-agnostic classifiers, where the training sets are broader.

For intent recognition, we find in Table \ref{tab:ResultsTerminalSpecificRecog} that the distance between the user and the terminal appears to correlate well with our results. As shown in Table \ref{tab:TerminalDetails}, Terminals 2 and 6 are an arm's length away from the user (and yield weak results), Terminals 1, 4, and 5 are near, and Terminal 3 protrudes. Here, the protrusiveness of Terminal 3 works against us, as the smoother gesture that results from the change in pose is less distinctive; whereas, for Terminals 1, 4, and 5, no such change in pose was prompted, so users interacted with these terminals with a contorted arm twist, causing a more conspicuous and distinct gesture. Terminal 1 proved to be particularly awkward for shorter users, causing the most conspicuous gesture in that case. We find that user comfort is beneficial in authentication, but detrimental in intent recognition.

The strongest results in Table \ref{tab:ResultsTerminalSpecificRecog} are given by Terminal 1 and the weakest by Terminals 2, 3, and 6. The former is flat on the surface and so demands the greatest wrist rotation from the user, whereas the latter three are inclined at angles of $45^{\circ}$ or greater (see Table \ref{tab:TerminalDetails}) and so require the least. This echoes our finding in Section \ref{sec:ResultsFeatureInformativeness} and suggests that wrist rotation is a key discriminator between tap gestures and other gestures.

The results for the freestyle terminal improve significantly with larger windows. Here, the freedom to manipulate both the smartwatch and the terminal leads not only to a smoother gesture, but also to a shorter alignment phase, both of which likely contribute to weaker results; however, for $s=3$, the results are better than those for the agnostic model, suggesting that the preparatory movements made in the reaching phase by users when interacting with the freestyle terminal, which included lifting the arm across the chest, are highly distinctive.

The terminal-agnostic approach is clearly superior to the terminal-specific approach for intent recognition, for all but the most awkwardly-positioned terminals. This shows that a classifier trained on tap gestures from a broader range of terminals becomes more effective at distinguishing a tap gesture from other gestures.

\begin{table}[t!]
	\centering
	\small
	\begin{tabular}{l|c|c|c}
		\toprule
		\textbf{Activity} & \textbf{Number of} & \textbf{Proportion of} & \\
		\textbf{Type} & \textbf{Samples} & \textbf{Samples (\%)} & \textbf{FAR (\%)} \\
		\midrule
		walking & 17890 & 55.15 & 2.99 \\
		bus or train & 9463 & 29.17 & 4.66 \\
		in-store & 4417 & 13.62 & 6.08 \\
		\midrule
		all & 32441 & 100 & 3.77 \\
		\bottomrule
	\end{tabular}
	\caption{Average FARs (tuned to the EER) by non-tap gesture type in optimum window $\{s=2.5, o=0\}$ for our intent recognition model. This excludes combined activity data, which account for 2\% of non-tap gesture samples.}
	\label{tab:ResultsFARBySetting}
\end{table}

\subsection{Enrolment Parameters}\label{sec:sec:ResultsEnrolmentParameters}

\begin{figure}[t!]
	\centering
  	\includegraphics[width=0.76\linewidth]{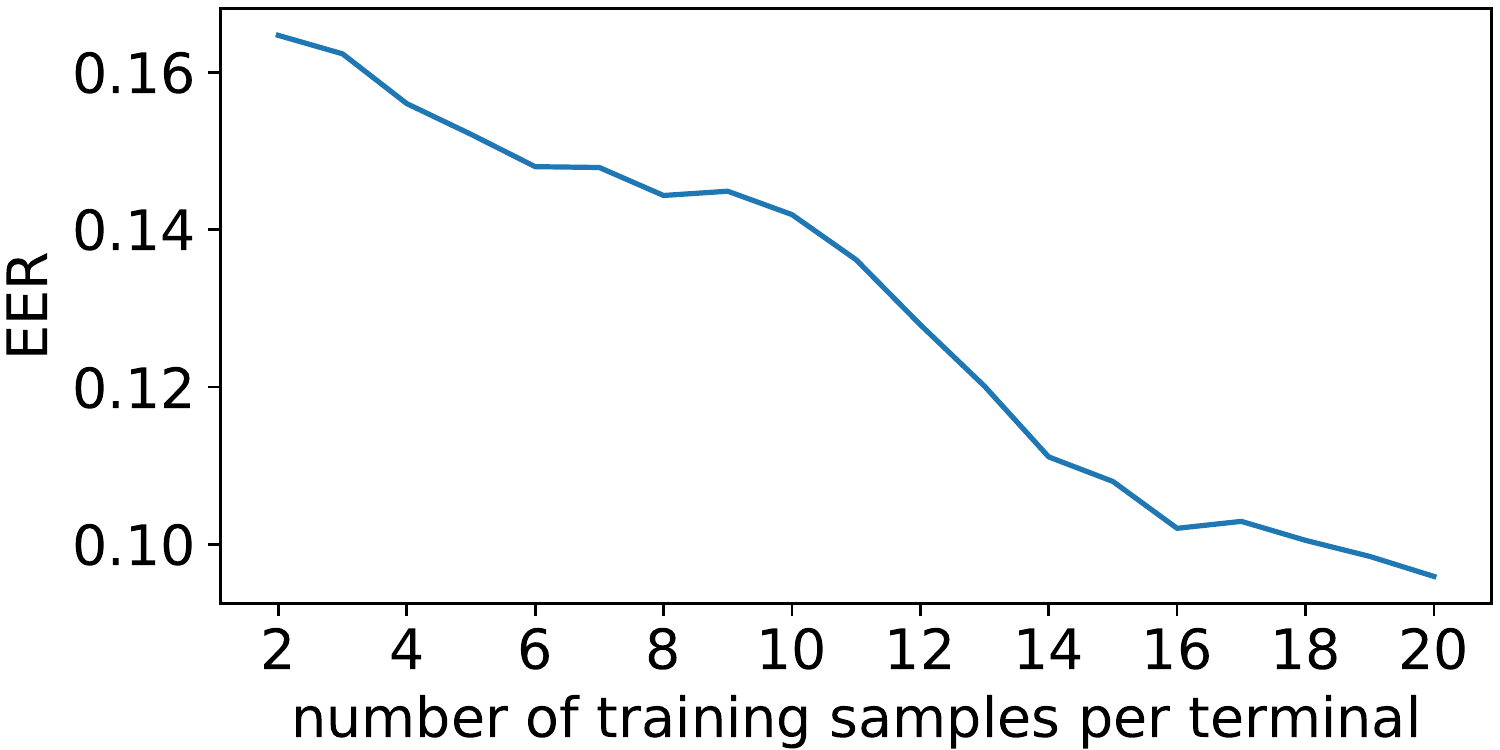}
   	\caption{Average EERs for our authentication model if trained on different numbers of enrolment samples in optimum window $\{s=2.5, o=0\}$. Each classifier is trained on six terminals.}
   	\label{fig:ResultsTrainingSize}
\end{figure}

Behavioural biometric systems typically entail a burdensome enrolment phase, where the user must perform the measured characteristic repeatedly to create the initial template. To evaluate the extent to which we can expedite the enrolment phase, we compare the average EERs of our authentication model when the classifiers are trained on smaller positive classes (\textit{i.e.}, fewer user samples). Figure \ref{fig:ResultsTrainingSize} shows that our model can authenticate the user with an average EER of 0.16 when it is trained on just 12 of the user's tap gestures (spread evenly over six terminals), which can be performed in less than a minute. We see that the EER improves as more samples are included in the training set; this suggests that an update mechanism might benefit the model over time, relaxing upfront requirements and incorporating subsequent tap gestures as the system is used. (Note that our intent recognition model is user-agnostic, so does not require training data from the user.)

\subsection{Misclassification by Activity Type}

To see where misclassifications in our intent recognition model are most likely to occur, we sort them by activity type. Table \ref{tab:ResultsFARBySetting} shows the number, proportion, and FAR of gesture samples of each type. We see that in-store gestures account for the greatest proportion of false positives; it is likely that actions such as reaching for a product on a shelf and rotating the wrist to read the label on a product exhibit some similar movement pattern fragments as those found in a tap gesture.

\begin{figure}[t!]
	\centering
  	\includegraphics[width=0.60\linewidth]{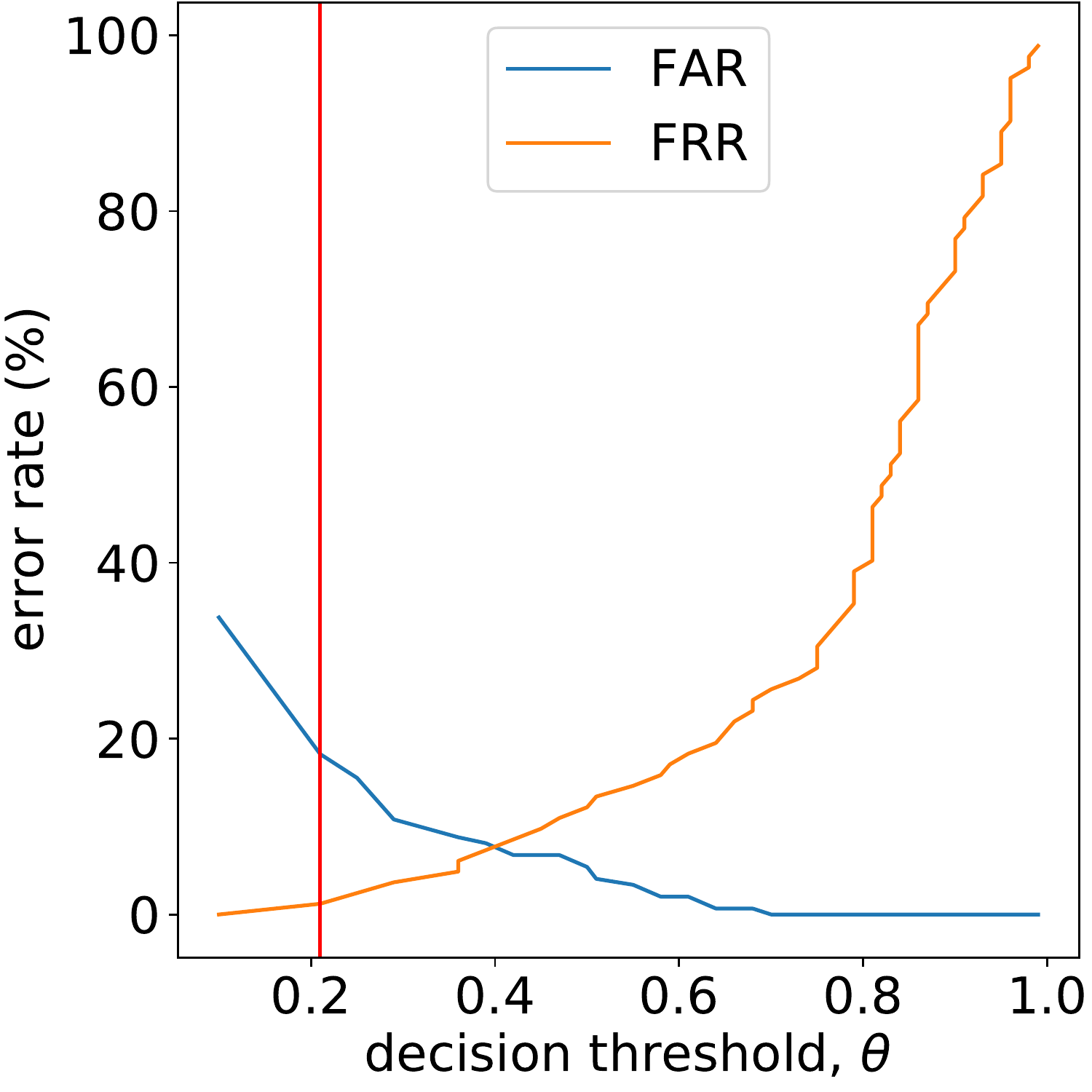}
   	\caption{An example plot showing an EER of 7\% at $\theta=0.42$. At $\theta=0.21$ (indicated with a red line), the model is optimised to minimise the FRR and has an FAR of 18\%.}
   	\label{fig:ResultsDecisionThreshold}
\end{figure}

\subsection{Cost of Optimisation}

For our models to add a strict improvement to the security of an existing system, we need to ensure that we do not impose a burden on its usability, so we optimise the classifiers to minimise the occurrence of false negatives that would impose inconvenience (see Figure \ref{fig:ResultsDecisionThreshold}). Figures \ref{fig:ResultsEERAuth} and \ref{fig:ResultsEERRecog} show the EERs of our models (\textit{cf.} Figure \ref{fig:ResultsF1}, which shows the F-measure scores). Figures \ref{fig:ResultsFARAuth} and \ref{fig:ResultsFARRecog} show the FARs of each classifier once its decision threshold has been adjusted to minimise the FRR. Figures \ref{fig:ResultsFARdeltaAuth} and \ref{fig:ResultsFARdeltaRecog} show the differences between the respective FARs before and after optimisation and therefore the cost.

We see that the cost of optimisation for authentication is relatively low; whereas, for intent recognition, the EERs start lower but the costs are much higher, suggesting steep FAR curves in those windows with fewer alignment phase samples. We find that the optimum window parameters for the optimised models for in-store usage are $\{3\le s\le4,\\o=0\}$, which conveys a cross-model average FAR of 0.15, and for historic analysis, $\{s=4, -2\le o\le-1\}$, with an average FAR of 0.13.

Our approach adds two components to the security of an existing system. Firstly, we add a layer of false acceptance detection, the effectiveness of which is shown by these low FARs. Secondly, we introduce to the system an unsharable factor that ensures that only the legitimate user can make payments with the smartwatch. (Note that unsharability is an oft overlooked yet advantageous property of biometrics and one that cannot otherwise be achieved by knowledge- or possession-based factors.) The results above have shown that we are able to provide these security gains without imposing a burden on usability.

\section{Discussion}\label{sec:Discussion}

\textbf{Power Consumption.} Smartwatches are designed to facilitate always-on sensing (\textit{e.g.}, in health and fitness monitoring). To measure the impact of our data collection app in practical terms, we wore two Samsung Galaxy Watches in an identical state, but with one running our app. Without any effort put into performance optimisation, our app caused the watch running it to consume an additional 1.5\% of battery capacity per hour. While we do not implement the random forest classifier on the smartwatch, we argue that its energy consumption would be negligible due to the limited number of inferences that would be required per day (only when the user makes a payment).

\textbf{Response Time.} We calculated the computation time for classifying a single tap gesture, averaged over 10,000, to be 7.11 ms for authentication and 7.09 ms for intent recognition on a desktop computer with an Intel Core i5-6500 processor. Using a benchmarking tool\footnote{https://www.notebookcheck.net}, we found that a Samsung Exynos W920 (a modern smartwatch processor) performs 26 times slower, so we would expect a response time of roughly 185 ms on a smartwatch for in-store usage. This could not be tested directly due to a lack of library support in Tizen Studio IDE.

\section{Related Work}\label{sec:RelatedWork}

\textbf{Authentication.} With regard to authentication, Shrestha \textit{et al.} \cite{Shrestha2016} present the most closely related work. They consider a system model in which the user makes mobile payments with a smart\textit{phone} and is authenticated by tap gesture. In the authentication portion of our work, we assume a similar context but explore the use of wrist-worn sensors, producing a physiologically distinct gesture and introducing a number of additional challenges (as described in Section \ref{sec:DesignConsiderations}). They achieve F-measure scores of up to 0.93 for authentication, a slight improvement on our results; however, they use cross-validation to train a classifier in an authentication use-case, which violates the requirement that training (enrolment) should precede testing (user verification) \cite{Eberz2017-1}, potentially inflating their scores. Drilling into the results, we note that their classifiers consistently had higher scores for recall than for precision---ours had the opposite, suggesting that the smartwatch gesture favours security and the smartphone, usability. They find the ideal gesture size to be 1 second of sensor data and mention losses in accuracy for greater sizes due to the capturing of extraneous movements---we find different optimum parameters for our wrist-led gesture (as described in Section \ref{sec:ResultsOptimumWindowParameters}); furthermore, our sliding window approach makes it possible for us also to consider the case of retrospective fraud detection. They collect data with terminals set in generic positions---we set ours in positions matching real-world terminals to elicit a truer representation of real payment gestures in our data; furthermore, we include a freestyle terminal to incorporate the common scenario of a vendor handing the terminal to the customer and to counter overfitting in our models. We tabulate these differences in Table \ref{tab:RelatedWorkComparisonShrestha}.

Lee \textit{et al.} \cite{Lee2017} use inertial sensors on a smartphone to authenticate the user whenever the phone is picked up, defining the implicit \textit{pick up gesture}, with the goal of reducing the number of explicit log-in actions required. Similar prior work by Conti \textit{et al.} \cite{Conti2011} authenticates the user as he makes or answers a phone call. Both works show solid results with short, simple gestures using a phone. These approaches use dynamic time warping to analyse sensor data; we instead use machine learning classifiers that expose the relative importances of the features upon which they base their decisions to refine our feature set and to observe the impact of our axis-invariant features.

Johnston \textit{et al.} \cite{Johnston2015} use inertial sensors on a smartwatch to infer gait as the user walks for identification and authentication purposes, using 10-second windows of sensor data.  Acar \textit{et al.} \cite{Acar2020} use inertial sensors on a smartwatch, in combination with keystroke dynamics measured at a workstation, to continuously authenticate the user against insider attacks when typing at a keyboard, achieving strong results with 20 seconds of sensor data. Lee \textit{et al.} \cite{Lee2016} consider the use of inertial sensors on a smartwatch (or other wearable device) as ancillary sensors to an authentication system based on a smartphone, although not in isolation. Orthogonal implicit authentication systems on smartwatches have adapted heart rate biometrics to authenticate the user using electrocardiography (ECG, electrical-based) or photoplethysmography (PPG, light-based) sensors \cite{Ekiz2020,Ramli2016,Vhaduri2019}. These systems require a few minutes to calibrate yet show promise over time, although ECG sensors have been shown to be vulnerable to spoofing attacks \cite{Eberz2017-2}.

Some works use inertial sensors on a smartwatch to authenticate the user with an explicit gesture, made solely for the purpose of authentication, such as MotionAuth \cite{Yang2015} (full arm gestures), ThumbUp \cite{Yu2020} (hand and finger gestures), and work by Liang \textit{et al.} \cite{Liang2017} (a punch gesture). The use of an explicit gesture can achieve strong results, but the user must spend time to perform it and must remember it, each of which can impose an inconvenience.

\begin{table}[t!]
	\centering
	\small
	\begin{tabular}{l|c|c}
		\toprule
		 & \textbf{This} & \\
		\textbf{Key Aspects} & \textbf{Work} & \textbf{ \cite{Shrestha2016} } \\
    	\midrule
        device used for tapping & watch & phone \\
        authentication & \checkmark & \checkmark \\
       	intent recognition & \checkmark & $\times$ \\
       	real-world inspired terminal set-up & \checkmark & $\times$ \\
       	inclusion of a non-fixed terminal & \checkmark & $\times$ \\
       	in-store usage/real-time use-case & \checkmark & \checkmark \\
       	retrospective fraud detection use-case & \checkmark & $\times$ \\
       	additional factor use-case & \checkmark & $\times$ \\
       	\bottomrule
	\end{tabular}
	\caption{Comparison of the key aspects of this work with those of the most closely related work,  Shrestha \textit{et al.} \cite{Shrestha2016}.}
	\label{tab:RelatedWorkComparisonShrestha}
\end{table}

\textbf{Intent Recognition.} With regard to intent recognition, we infer an intent-to-pay if we identify a tap gesture. This is a novel contribution and to the best of our knowledge there is no closely related work. Loosely, we know of two works that infer a security feature from wrist-based activity recognition: Mare \textit{et al.} present both ZEBRA \cite{Mare2014} and CSAW \cite{Mare2019}, which infer activities from wrist-worn sensor data and correlate them, respectively, with workstation inputs or motion sensor data from a smartphone to (de-)authenticate continued usage by the user. These systems, like the intent recognition portion of our work, achieve their inferences in a user-agnostic manner.

\section{Limitations and Future Work}\label{sec:LimitationsandFutureWork}

The main limitation of this work is the size of the dataset (unfortunately, our experimental work was stopped abruptly by national lockdowns in 2020). Having samples from 16 users enables us to demonstrate the feasibility of our approach; however, to validate our findings, more users are required and this should be a focus of future work. The collection of tap gestures in an artificial lab setting, notwithstanding our efforts to immerse the user, is also a limitation; future work should gather tap gestures from payments made in the wild to ensure that the system is robust against noise caused by real-world obstacles and distractions that affect the user.

\section{Conclusion}\label{sec:Conclusion}

In this paper, we showed that a tap gesture can be used to authenticate the user and recognise intent-to-pay, implicitly, while the user makes a payment with a smartwatch. Our approach is software-driven and does not require any changes to terminals. Our authentication model is terminal-agnostic, so does not require the use of any specific terminal type or position, and achieves F-measure scores of up to 0.87 and EERs as low as 0.08. Our intent recognition model is user-agnostic, so does not require the user to provide any training data during enrolment and is resistant to drift, and achieves F-measure scores of up to 0.89 and EERs as low as 0.04. We found the optimum gesture parameters for in-store usage and for retrospective fraud detection. We showed that our models can be optimised for usability and incorporated as an additional factor in an existing system to provide a strict improvement to security (in terms of FAR and by adding an unsharable factor) at negligible cost to usability (in terms of FRR). We identified that the tap gesture is triphasic and analysed how this and other factors contributed to our results. Finally, we explored the applicability of our approach to alternative system models with fewer input sensors, dedicated terminals, or relaxed enrolment requirements while remaining performant.

Without loss of generality, we have focused on the context of mobile payments. Our approach has wide applicability to any user authentication context in which a task or gesture is performed while wearing a smartwatch, such as building access control, vehicle access control, or interaction with smart devices or objects.

\section*{Acknowledgement}

This work was supported financially by Mastercard; the Engineering and Physical Sciences Research Council [grant number EP/P00881X/1]; and the PETRAS National Centre of Excellence for IoT Systems Cybersecurity [grant number EP/S035362/1]. The authors would like to thank these organisations for their support and the anonymous reviewers for their feedback.

\balance

\end{document}